\documentclass[twocolumn,english,aip,reprint]{revtex4-1}
\usepackage[T1]{fontenc}
\usepackage[latin9]{inputenc}
\usepackage{babel}
\usepackage{amsmath}
\usepackage{amssymb}
\usepackage{graphicx}
\usepackage{wasysym}
\usepackage[unicode=true]
 {hyperref}


\begin{document}

\title{Border-collision bifurcations in a driven time-delay system}

\author{Pierce Ryan}

\affiliation{School of Mathematical Sciences, University College Cork, Cork T12
XF62, Ireland}

\author{Andrew Keane}

\affiliation{Department of Mathematics, The University of Auckland, Private Bag
92019, Auckland 1142, New Zealand}

\affiliation{School of Mathematical Sciences, University College Cork, Cork T12
XF62, Ireland}

\author{Andreas Amann}

\affiliation{School of Mathematical Sciences, University College Cork, Cork T12
XF62, Ireland}
\begin{abstract}
We show that a simple piecewise-linear system with time delay and
periodic forcing gives rise to a rich bifurcation structure of torus
bifurcations and Arnold tongues, as well as multistability across
a significant portion of the parameter space. The simplicity of our
model enables us to study the dynamical features analytically. Specifically,
these features are explained in terms of border-collision bifurcations
of an associated Poincaré map. Given that time delay and periodic
forcing are common ingredients in mathematical models, this analysis
provides widely applicable insight. 
\end{abstract}
\maketitle
\begin{quotation}
Both time-delay dynamical systems, and periodically driven dynamical
systems have been thoroughly studied in the literature. This can be
attributed to their great relevance to real-world problems. Time-delay
systems arise naturally in physical, biological or climate models
due to finite propagation speed; periodic drive is ubiquitous in engineering
applications and is known to generate complex resonance phenomena.
However, systems that combine these two properties have received much
less attention, despite being relevant in many real-world applications.
In this paper, we study a simple piecewise-linear system with both
time delay and periodic forcing, which exhibits interesting dynamical
features as a nontrivial consequence of this combination. These features
include multistabilities, Arnold tongues and torus bifurcations. Since
the system is piecewise linear and contains only two parameters, many
phenomena can be interpreted through an analytically derived piecewise-smooth
Poincaré map and an analysis of the associated border-collision bifurcations.
The analysis explains the origin of similar phenomena which has previously
been observed numerically in more complicated related systems.

This article may be downloaded for personal use only. Any other use
requires prior permission of the author and AIP Publishing. This article
appeared in Chaos 30, 023121 (2020) and may be found at \href{https://aip.scitation.org/doi/10.1063/1.5119982}{https://aip.scitation.org/doi/10.1063/1.5119982}.
\end{quotation}

\section{Introduction}

Periodically driven systems appear in many real-world applications.
Examples include optical injection in laser systems \citep{wieczorek_dynamical_2005},
vibration-driven energy harvesting devices \citep{beeby_micro_2007},
injection-locked frequency dividers in electronics \citep{daneshgar_observations_2010},
or seasonal forcing in climate systems \citep{tziperman_nino_1994}.
They often give rise to interesting resonance behaviour in damped
oscillators \citep{parlitz_superstructure_1985} and complex synchronization
patterns in self-sustained oscillators \citep{boccaletti_synchronization_2018,marchionne_synchronisation_2018}.

Similarly, time-delay systems also arise in many experimental systems,
for example in optics \citep{heil_chaos_2001}, electronics \citep{larger_virtual_2013},
neuro-science \citep{scholl_time_2009}, or climate systems \citep{runge_quantifying_2014},
and also play an important role in chaos control, for example, through
the use of time-delayed feedback control \citep{pyragas_control_1995}.
From a mathematical point of view, time delay often leads to a formally
infinite-dimensional phase space \citep{hale_introduction_2013},
which considerably complicates the analysis, but allows for a rich
variety of phenomena.

The combination of external forcing and time delay has been studied,
for example, in the context of the Duffing oscillator \citep{hu_resonances_1998},
the van der Pol oscillator \citep{maccari_vibration_2003} and more
recently in the context of climate systems \citep{ghil_delay_2008,keane_delayed_2015}.
However, a general understanding of this class of dynamical systems
is not yet available. The objective of the current paper is, therefore,
to study the fundamental features of an elementary system with time
delay and periodic forcing to obtain a broader insight into what phenomena
are expected to arise as a consequence of this combination.

Let us consider a simple driven time-delay dynamical system introduced
by Ghil et al. \citep{ghil_delay_2008} as a model for a climate phenomenon
known as the El Niño Southern Oscillation. The system of a real variable
$x\in\mathbb{R}$ is defined by
\begin{align}
\dot{x}(t) & =-\text{\ensuremath{\tanh}}\left[\kappa x\left(t-\tau\right)\right]+b\sin\left(2\pi t\right),\label{eq:1}\\
x(t) & \in C\left(\left[-\tau,0\right]\right),\label{eq:2}
\end{align}
where $b\geq0$ is the magnitude of the periodic forcing, $\tau>0$
is the time delay of the delayed feedback, and $\kappa>0$ is the
linear slope of the delayed feedback at the origin. A solution of
the system is a trajectory $x(t)\in\mathbb{R}$, $t\in\mathbb{R}$.
A consequence of the reliance of the delayed feedback on a continuous
function $x\left(t\right)$ over an interval $\left[-\tau,0\right]$
is that the system has an infinite-dimensional phase-space. Another
key feature of this system is that it has the symmetry $x\left(t\right)\rightarrow-x\left(t+\frac{1}{2}\right)$.
This model has been studied extensively by Keane et al. \citep{keane_delayed_2015,keane_investigating_2016}.
Numerical analysis of this system demonstrated an extremely complex
resonance structure\citep{keane_delayed_2015}. Furthermore, the autonomous
system $\left(b=0\right)$ has been studied analytically\citep{nussbaum_uniqueness_1979,chow_characteristic_1988}.
In this case the trivial solution $x\equiv0$ is only stable for $\tau<\frac{\pi}{2\kappa}$.
At $\tau=\frac{\pi}{2\kappa}$, it becomes unstable, and a family
of stable $4\tau$-periodic solutions is born. 

In order to analyse the phenomena seen in this model further, let
us consider a further simplification of the system by taking $\kappa\rightarrow\infty$.
This has the effect of changing the delayed feedback term from $-\text{\ensuremath{\tanh}}\left[\kappa x\left(t-\tau\right)\right]$
to $-\text{\text{sgn}}\left[x\left(t-\tau\right)\right]$. We also
apply the signum function to the periodic forcing to obtain the dynamical
system
\begin{align}
\dot{x}(t) & =-\text{sgn}\left[x\left(t-\tau\right)\right]+b\text{ sgn}\left(\sin\left(2\pi t\right)\right),\label{eq:3}\\
x(t) & \in C\left(\left[-\tau,0\right]\right),\label{eq:4}
\end{align}
where $b\geq0$ and $\tau>0$. Critically, this simplification of
the system preserves the symmetry $x\left(t\right)\rightarrow-x\left(t+\frac{1}{2}\right)$.
The feedback term $-\text{ sgn}\left[x\left(t-\tau\right)\right]$
takes values in $\left\{ 1,0,-1\right\} $. However, while the feedback
can in principle be $0$, this occurs only under highly specific conditions
which are not considered here. The forcing term $b\text{ sgn}\left[\sin\left(2\pi t\right)\right]$
takes values in $\left\{ b,0,-b\right\} $. As the forcing is $0$
only at discrete times, we say that the forcing is positive for $t\bmod1\in[0,0.5)$,
and negative for $t\bmod1\in[0.5,1)$. We now develop our method for
solving the system. 

\section{Numerics}

\subsection{Iterative map}

In order to solve Eqs.~(\ref{eq:3},\ref{eq:4}), we note that $\dot{x}(t)$
can only take discrete values in $\left\{ 1+b,-1+b,1-b,-1-b\right\} $;
therefore this continuous system can be modelled exactly as a discrete
time iterative map, or Poincaré map. The state of the system at time
$t$ is a tuple of variable length
\begin{equation}
S(t)=\left(x;z_{0},z_{1},...,z_{n-1}\right),
\end{equation}
where $x\in\mathbb{R}$ is the position at time $t$ and $t-\tau<z_{0}<z_{1}<...<z_{n-1}\leq t$
are \emph{zero elements}, which are the times at which the trajectory
passed through $x=0$ in $(t-\tau,t]$. Let $n$ be the number of
zero elements in $S(t)$. Thus, Eq.~(\ref{eq:3}) may be rewritten
as
\begin{equation}
\dot{x}(t)=-\text{sgn}\left[x\left(t\right)\right](-1)^{n}+b\text{sgn\ensuremath{\left[\sin\left(2\pi t\right)\right]}}.
\end{equation}
This representation of the state of the system shows that the dimension
of this system is finite, but variable. This will have a significant
impact in our analysis of this system.

There are three key events that occur in a trajectory that affect
$S(t)$:
\begin{enumerate}
\item A zero element is added. When $x(t)$ passes through $x=0$, a zero
element equal to $t$ is appended to $S$, increasing $n$ by $1$;
this does not immediately cause the feedback to change as the sign
of $x(t)$ also changes, but it will affect $\dot{x}$ at time $t+\tau$.
\item A zero element is removed. When $t=z_{0}+\tau$, $z_{0}$ is deleted
from $S$ which decreases $n$ by 1, resulting in a sign change of
the feedback.
\item The forcing changes when $t\bmod0.5=0$. While this does not affect
the state directly, it changes $\dot{x}$, affecting the evolution
of the state.
\end{enumerate}
Between consecutive events, $\dot{x}$ is constant. We construct our
iterative map to move the system forward in time to the next event.
The step size $\Delta t$ for the iterative map is variable. One iteration
of the map consists of calculating $\dot{x}(t)$, calculating the
step size $\Delta t$, then calculating $S\left(t+\Delta t\right)$.
This process is used to simulate the trajectory of the system from
any given initial state $S(0)$ up to a maximum time $T$.

\subsection{Sample solutions}

\begin{figure}[!t]
\includegraphics[width=1\columnwidth]{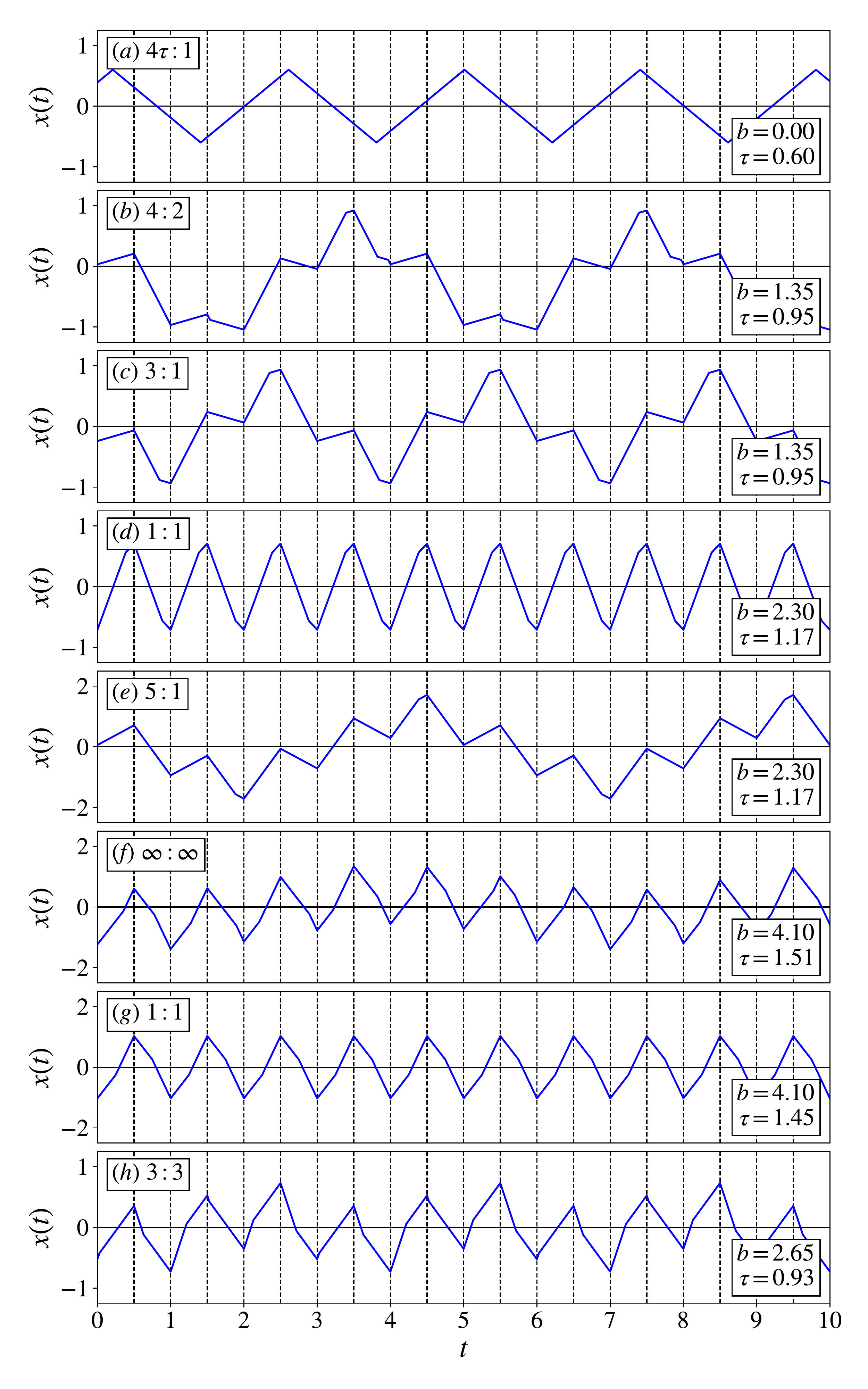}

\caption{\label{fig:sample_solutions}Sample solutions, excluding transients,
obtained from simulating the system from different initial conditions
and parameters $(b,\tau)$ shown in the bottom right corner of each
plot. The ratio of the period of the solution to the number of crossings
from $x<0$ to $x>0$ in one cycle is shown in the top left corner
of each plot. The vertical dotted lines indicate times when the forcing
changes. }
\end{figure}

We now present sample solutions which demonstrate some of the characteristic
features of this system. Fig.~\ref{fig:sample_solutions} shows eight
stable solutions found by simulating the system from different initial
conditions and parameters $(b,\tau)$. 

There are periodic and aperiodic solutions present in the system.
A periodic solution with period $P$ is a trajectory that follows
a cycle such that $S(t+P)=S(t)$. An aperiodic solution is considered
to be a solution with infinite period. We label solutions by the \emph{characteristic
ratio} $P\mathbin{:}R$, where $R$ is the number of times the trajectory
crosses from $x<0$ to $x>0$ in one cycle. Note that the $P\mathbin{:}R$
notation differs from the $p\mathbin{:}q$ notation used in some previous
literature\citep{keane_delayed_2015,keane_investigating_2016}, where
$\frac{p}{q}$ is the rotation number. We find that $P\mathbin{:}R$
is a useful measure of a solution, for reasons that will be made clear
later.

Fig.~\ref{fig:sample_solutions}(a) is an example of the stable solution
to the unforced system $\left(b=0\right)$. The change in the feedback
occurs at a time $t+\tau$ after the trajectory passes through $x=0$
at time $t$, resulting in a $4\tau$-periodic solution that is stable
for $\tau>0$. This is consistent with the analytic results found
for the unsimplified system. We consider $4\tau$ to be the natural
period of the feedback, as the solution to the unforced system is
$4\tau$-periodic. The characteristic ratio of this solution is $4\tau\mathbin{:}1$.

Fig.~\ref{fig:sample_solutions}(b) and Fig.~\ref{fig:sample_solutions}(c)
show a $4\mathbin{:}2$ solution and a $3\mathbin{:}1$ solution,
respectively, that are stable for the same parameters. This is an
example of bistability, where there are two stable solutions for the
same parameters; the solution to which the system converges depends
on which solution's basin of attraction the initial conditions are
in. A second example of bistability is seen in Fig.~\ref{fig:sample_solutions}(d)
and Fig.~\ref{fig:sample_solutions}(e), which show a $1\mathbin{:}1$
solution and a $5\mathbin{:}1$ solution, respectively, that are stable
for the same parameters. 

The solution in Fig.~\ref{fig:sample_solutions}(f) is assumed to
be aperiodic, as the system does not converge to a periodic solution
after running a simulation up to $T=100000$. By comparing the aperiodic
solution to the $1\mathbin{:}1$ solution in Fig.~\ref{fig:sample_solutions}(g),
we may note that the aperiodic solution and the $1\mathbin{:}1$ solution
have the same number of $x=0$ crossings per period of the forcing.
However, the $x=0$ crossings are evenly spaced in the $1\mathbin{:}1$
solution, but are not in the aperiodic solution. This may be an indicator
that the aperiodic solution is related to the $1\mathbin{:}1$ solution
in Fig.~\ref{fig:sample_solutions}(g). A similar observation can
be made for the $3\mathbin{:}3$ solution in Fig.~\ref{fig:sample_solutions}(h)
and the $1\mathbin{:}1$ solution in Fig.~\ref{fig:sample_solutions}(d).
In later sections of this paper, we will investigate these relationships
in detail.

\subsection{Structure in the $\left(b,\tau\right)$ plane}

\begin{figure}[!t]
\includegraphics[width=1\columnwidth]{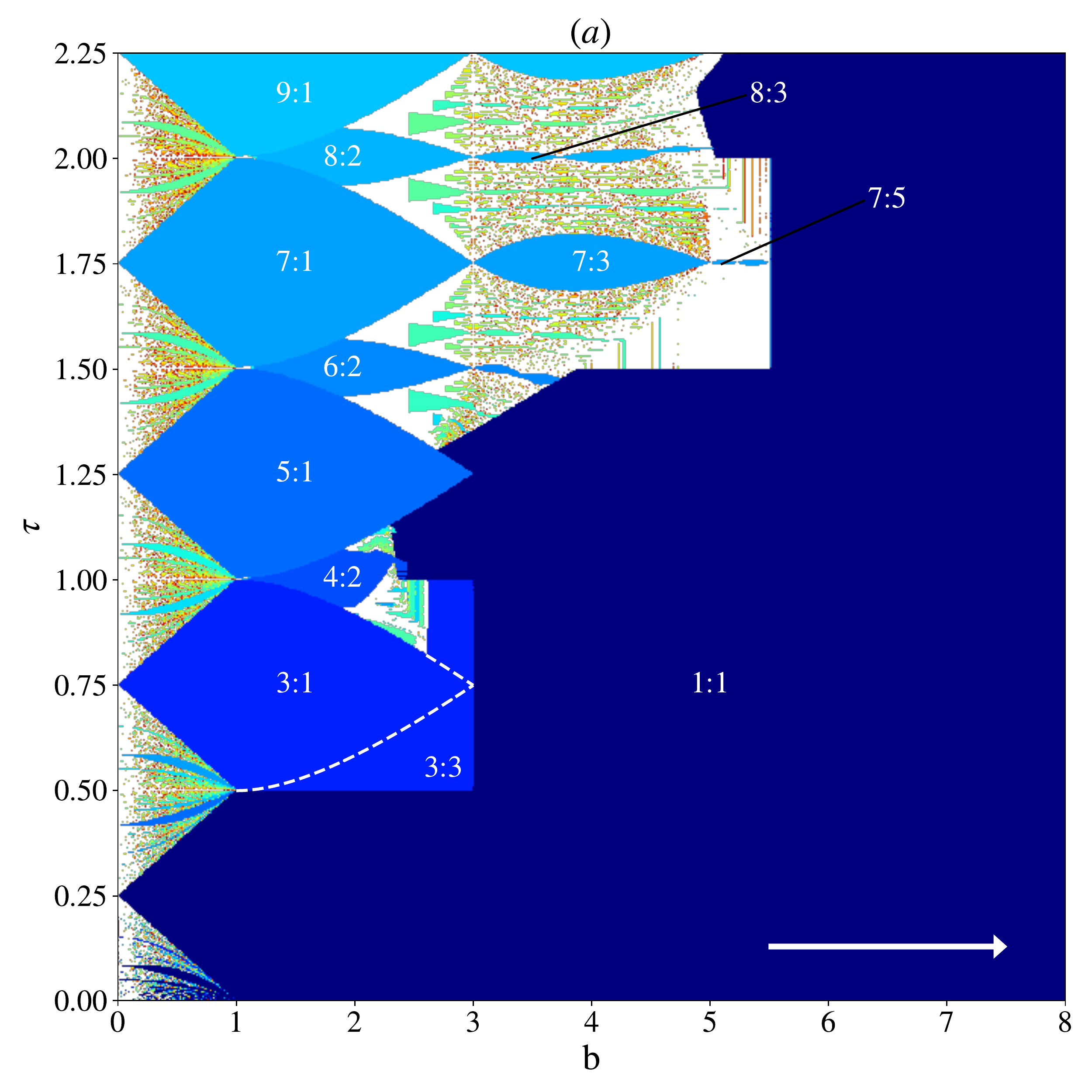}

\includegraphics[width=1\columnwidth]{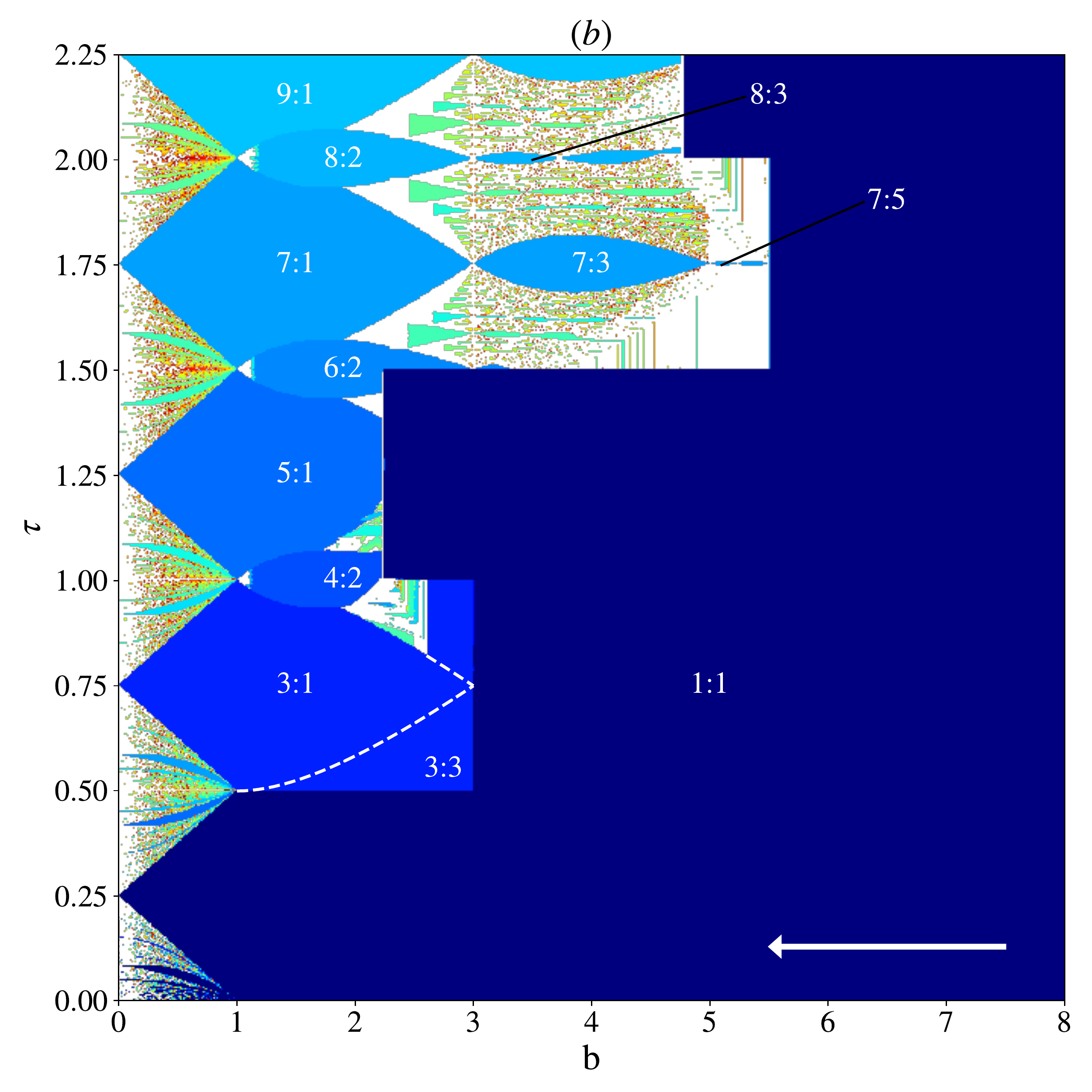}

\caption{\label{fig:period}Colour maps showing the period of simulated solutions
obtained by making 1124 sweeps of 1125 simulations of duration $T=10000$
in $[0,8]\times[0,2.5]$. The white arrows indicate the direction
of the sweep. Dark blue indicates $P=1$ solutions, shading to dark
red for $P=999$ solutions. White space indicates where the solution
was aperiodic, or had $P>999$. The white text indicates the characteristic
ratio of stable solutions found within the tongues. The dashed white
line shows the border between the regions where the $3\mathbin{:}1$
and $3\mathbin{:}3$ solutions occur.}
\end{figure}

Having seen some interesting features of the system, we move on to
understanding the overall dynamics in the $(b,\tau)$ plane. Due to
the bistability present in the system, simulating the system from
arbitrary initial conditions across a $(b,\tau)$ mesh would produce
an inconsistent picture, as we have no prior knowledge of the basins
of attraction of bistable solutions. In order to circumvent this issue,
for fixed $\tau$, the solution is swept across a range of $b$ by
iteratively simulating the system, incrementing $b$ slightly, then
simulating again using the final state of the previous simulation
as the initial state of the next one. This allows a stable solution
to be followed until it loses stability or ceases to exist, at which
point the system converges to a nearby stable solution. By taking
multiple sweeps in $b$ for a range of fixed $\tau$ values, we obtain
the $P\mathbin{:}R$ charts shown in Fig.~\ref{fig:period}. Fig.~\ref{fig:period}(a)
shows the $P\mathbin{:}R$ chart under an upward sweep in $b$, from
left to right. Fig.~\ref{fig:period}(b) shows the $P\mathbin{:}R$
chart under a downward sweep in $b$, from right to left. This figure
demonstrates many striking features of the system which will be explored
in more detail. 

First, let us focus our attention on the upward $b$ sweep in Fig.~\ref{fig:period}(a).
We observe that there are regions in the $\left(b,\tau\right)$ plane
in which particular $P\mathbin{:}R$ solutions exist, such as the
labelled $3\mathbin{:}1$, $5\mathbin{:}1$, $7\mathbin{:}1$ and
$9\mathbin{:}1$ regions on the left side of the chart. These regions
are sections of Arnold tongues. An Arnold tongue is a region of the
$\left(b,\tau\right)$ plane, rooted on $b=0$, within which the feedback
and the forcing synchronise to produce a solution with period equal
to a rational ratio of the forcing. In piecewise-linear systems, an
Arnold tongue can have shrinking points, at which the Arnold tongue
has zero width. This phenomenon was first observed in the circle map\citep{wei_arnold_1987},
and analysed in detail in the context of piecewise-linear continuous
maps with single switching mechanisms\citep{simpson_shrinking_2009,simpson_structure_2016,simpson_structure_2018}.
This produces Arnold tongues that have been compared to strings of
sausages\citep{wei_arnold_1987}. We will refer to an individual ``sausage''
as a \emph{tongue}, and refer to a full ``string'' as an \emph{Arnold
tongue}. A notable feature of these Arnold tongues is that the characteristic
ratio is different in each tongue in the string. For example, the
$P=7$ Arnold tongue is rooted on $b=0$ at $\tau=1.75$, and consists
of the $7\mathbin{:}1$ tongue, the $7\mathbin{:}3$ tongue, the $7\mathbin{:}5$
tongue, and the unlabelled $7\mathbin{:}7$ tongue. In each Arnold
tongue, the leftmost tongue is a $P\mathbin{:}R$ tongue that is rooted
on $b=0$ at $\tau=\frac{P}{4R}$. $P$ is the same in every tongue
in the chain, but $R$ increases the further right the tongue is in
the string. 

Tongues with similar characteristic ratios tend to have similar shape,
with some variation. For example, the $6\mathbin{:}2$ and $8\mathbin{:}2$
tongues have identical shape; the $4\mathbin{:}2$ tongue is different.
The large $P\mathbin{:}1$ tongues noted earlier have identical shape
for $P>1$. The $1\mathbin{:}1$ Arnold tongue is unlike any other
Arnold tongue in shape. For large $b$, the forcing dominates the
feedback, which results in the system converging to the $1\mathbin{:}1$
solution exclusively. We refer to the region in which $b$ is large
enough as the locked region, where the system is locked to the period
of the forcing. The boundary of the locked region is unusual, being
made up of straight lines which meet at right angles. Branching off
from the horizontal lines of the boundary, there are vertical stripes
in which $P\mathbin{:}P$ solutions exist. We will devote considerable
attention to studying the $1\mathbin{:}1$ solution later.

Now we compare the upward $b$ sweep in Fig.~\ref{fig:period}(a)
to the downward $b$ sweep in Fig.~\ref{fig:period}(b). The most
significant difference occurs around $\left(b,\tau\right)=\left(2.5,1.25\right)$.
In the upward sweep, the system follows the $5\mathbin{:}1$ solution
to the edge of the $5\mathbin{:}1$ tongue, and there is a complicated
region of smaller tongues and aperiodicity above the $5\mathbin{:}1$
tongue. In the downward sweep, the system instead continues to follow
the $1\mathbin{:}1$ solution into the region where different features
occurred in the upward sweep. This agrees with the bistability of
the $1\mathbin{:}1$ and $5\mathbin{:}1$ solutions seen in Fig.~\ref{fig:sample_solutions}(d,e).
A less obvious difference is the bistability to the right of $b=1$;
note the apparent difference in shape of the $P\mathbin{:}2$ and
$P\mathbin{:}1$ tongues near this line. This occurs because the $P\mathbin{:}1$
solutions overlap with the $P\mathbin{:}2$ solutions, in agreement
with the bistability of the $4\mathbin{:}2$ and $3\mathbin{:}1$
solutions seen in Fig.~\ref{fig:sample_solutions}(b,c). The absence
of $P\mathbin{:}1$  tongues for even $P$ is notable. We numerically
observe such solutions in simulations, but only for small $b\apprle0.5$
and exactly $\tau=\frac{P}{4}$. 

\section{Dynamics}

\begin{figure}[!t]
\includegraphics[width=1\columnwidth]{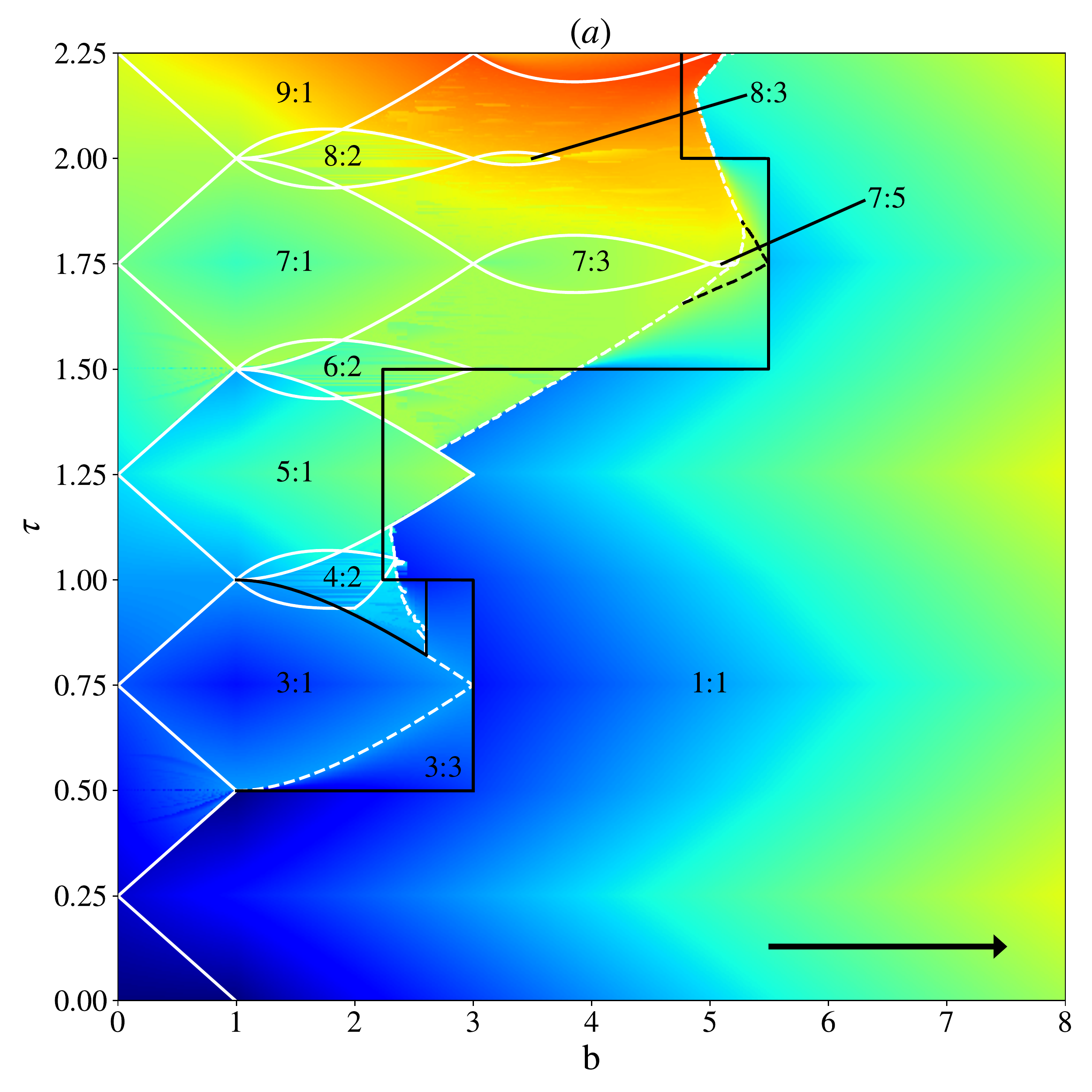}

\includegraphics[width=1\columnwidth]{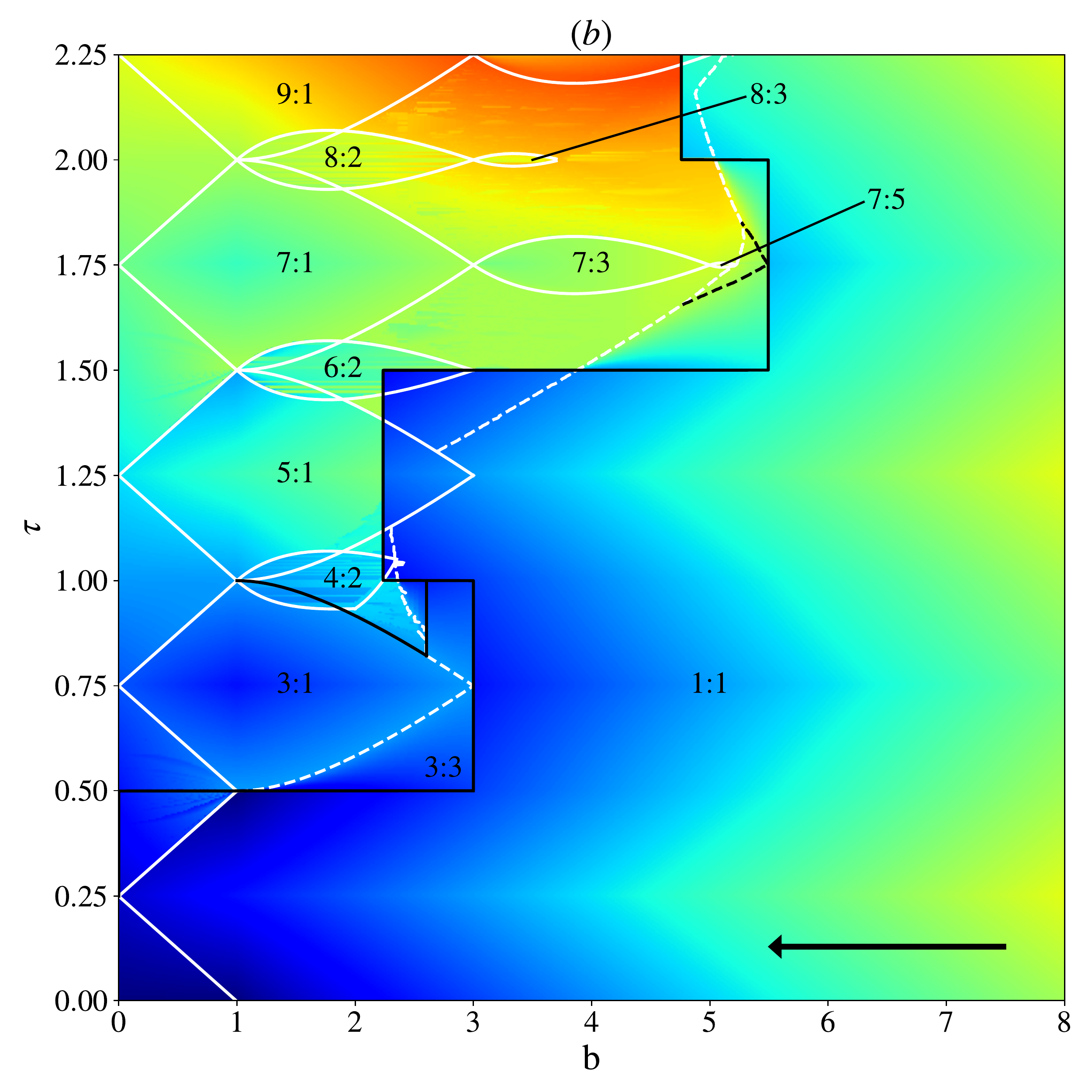}

\caption{\label{fig:bifurcations} Maximum charts in the $(b,\tau)$ plane
overlayed with bifurcation curves. Blue indicates a low maximum, red
indicates a high maximum. The black arrows indicate the direction
in which solutions were swept. The black text indicates the ratio
of the period of the solution to the number of $Z$ symbols in the
sequence for the stable solution within the associated tongue. BCSN
bifurcations are shown in solid white, and T bifurcations are shown
in solid black. The $D\bar{D}$ curve is shown in dashed white.}
\end{figure}

Fig.~\ref{fig:bifurcations} shows maximum charts in the $\left(b,\tau\right)$
plane overlayed with bifurcation curves. The maximum is taken as the
maximum value of a solution over an interval of length $1000$ after
a transient of length $9000$ from the same simulations that were
used to generate Fig.~\ref{fig:period}. Some of the larger tongues
seen in Fig.~\ref{fig:period} can be seen without difficulty in
Fig.~\ref{fig:bifurcations} due to a large jump in maxima at the
boundaries of the tongues. We note that the even $P\mathbin{:}R$
tongues are striped horizontally; this is most evident in the $4\mathbin{:}2$
and $8\mathbin{:}2$ tongues. It appears that solutions with even
$P$ or $R$ are not invariant under the symmetry $x\left(t\right)\rightarrow-x\left(t+\frac{1}{2}\right)$;
rather there exists a pair of symmetry-related counterpart solutions,
each with a different maximum value. The system converges to one of
these solutions depending on initial conditions, and remains at that
solution until swept out of the tongue, resulting in stripes parallel
to the direction of the sweep. This feature was also observed in the
smooth system (\ref{eq:1},\ref{eq:2}) by \citet{keane_delayed_2015}.

We will devote the rest of this section to deriving and characterising
the bifurcation curves plotted in Fig.~\ref{fig:bifurcations}. In
order to do so, we first need to establish a systematic method of
analysing solutions. We apply this method to develop Poincaré maps
and border collision maps through which we study the bifurcations
present in this system.

\subsection{Symbolic representation}

We require a robust framework under which we can analyse the dynamics
of this system. We note that the characteristic ratio does not distinguish
between the two different forms of the $1\mathbin{:}1$ solution shown
in Fig.~\ref{fig:sample_solutions}(d,g). Observing the order in
which the feedback and forcing change after the trajectory passes
through $x=0$, we note that the feedback changes before the forcing
in Fig.~\ref{fig:sample_solutions}(d), and after the forcing in
Fig.~\ref{fig:sample_solutions}(g). In order to precisely capture
these differences, a more explicit labelling system is required. We
therefore adopt a symbolic representation of solutions. Symbolic representations
have previously been used to great effect in the study of iterative
maps\citep{simpson_structure_2016,metropolis_finite_1973}.

\begin{figure}[!t]
\includegraphics[width=1\columnwidth]{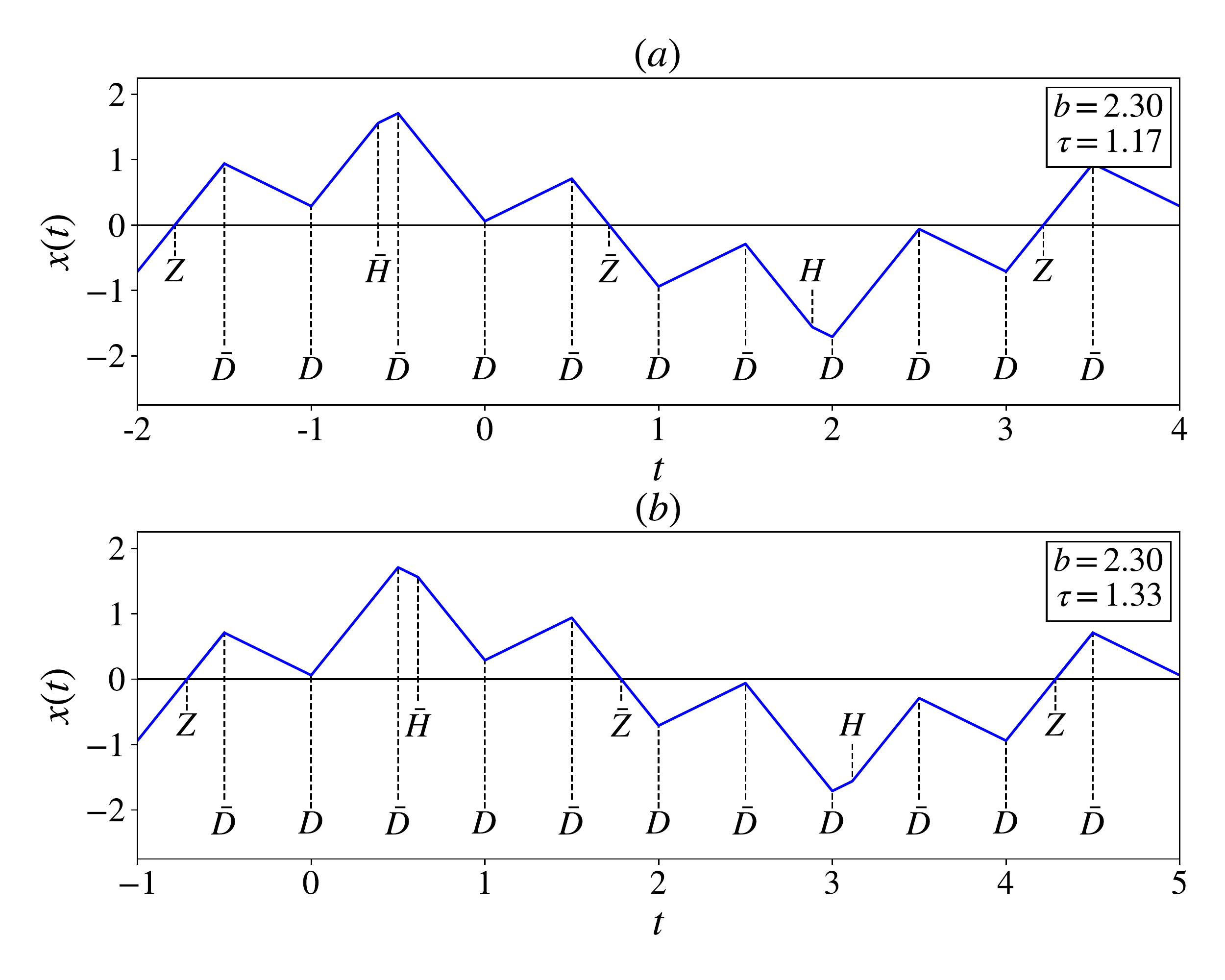}

\caption{\label{fig:symbolic} Derivation of the symbolic representations of
the $5\mathbin{:}1$ solution.}
\end{figure}

Let a solution be represented by a sequence of events $...X_{1}X_{2}...X_{n}...$
where $X_{i}\in\left\{ D,\bar{D},Z,\bar{Z},H,\bar{H}\right\} .$ $D$
denotes a transition of the forcing from $-b$ to $b$, $Z$ denotes
a transition from $x<0$ to $x>0$, and $H$ denotes a transition
of the feedback from $-1$ to $1$. A bar over a symbol causes it
to denote the opposite transition; a symbol with two bars over it
is the same as the symbol unbarred. Fig.~\ref{fig:symbolic}(a) shows
the $5\mathbin{:}1$ solution seen Fig.~\ref{fig:sample_solutions}(e),
labelled with the events that occur in the trajectory. As this solution
is periodic, the sequence repeats, so we abbreviate the sequence of
events representing the solution to a minimal repeating sequence $\left[Z,\bar{D},D,\bar{H},\bar{D},D,\bar{D},\bar{Z},D,\bar{D},H,D,\bar{D},D\right]$.
In general, a $P\mathbin{:}R$ solution is represented by a minimal
repeating sequence of events $\left[X_{1},X_{2},...,X_{n}\right]$
containing:
\begin{itemize}
\item $P$ $D$ events and $P$ $\bar{D}$ events,
\item $R$ $Z$ events and $R$ $\bar{H}$ events,
\item $R$ $\bar{Z}$ events and $R$ $H$ events.
\end{itemize}
Every cyclic permutation of a sequence represents the same solution.
For the sake of consistency, all sequences begins with a $Z$. We
observe that the $5\mathbin{:}1$ solution shown in Fig.~\ref{fig:symbolic}(a)
is invariant under the symmetry $x\left(t\right)\rightarrow-x\left(t+\frac{1}{2}\right)$;
this causes the second half of the sequence $\left[Z,\bar{D},D,\bar{H},\bar{D},D,\bar{D},\bar{Z},D,\bar{D},H,D,\bar{D},D\right]$
to be the same as the first half with all symbols barred. We abbreviate
$\left[Z,\bar{D},D,\bar{H},\bar{D},D,\bar{D},\bar{Z},D,\bar{D},H,D,\bar{D},D\right]$
as $\left[Z,\bar{D},D,\bar{H},\bar{D},D,\bar{D}\right]^{-}$ for convenience.
In general, a $P\mathbin{:}R$ solution of the form $\left[X_{1},X_{2},...X_{n},\bar{X_{1}},\bar{X_{2}},...,\bar{X_{n}}\right]$
can also be represented by a half-sequence $\left[X_{1},X_{2},...X_{n}\right]^{-}$.

A sequence is considered \emph{legal }within a subset of the $(b,\tau)$
plane if it represents a solution that exists within that subset.
Each $P\mathbin{:}R$ solution is represented by a set of legal sequences,
each one existing in a unique subset of the $(b,\tau)$ plane. The
union of these subsets is the region in which the solution exists.
The $5\mathbin{:}1$ solution exists within the $5\mathbin{:}1$  tongue
shown in Fig.~\ref{fig:period}(a). It is represented by the half-sequences
$\left[Z,\bar{D},D,\bar{H},\bar{D},D,\bar{D}\right]^{-}$ for $\tau\leq1.25$
and $\left[Z,\bar{D},D,\bar{D},\bar{H},D,\bar{D}\right]^{-}$ for
$\tau\geq1.25$, as shown in Fig.~\ref{fig:symbolic}. At $\tau=1.25$,
the increasing time delay between $Z$ and $\bar{H}$ causes $\bar{H}$
to swap with $\bar{D}$, changing the sequence. Determining whether
a given sequence is legal within a subset of the $(b,\tau)$ plane
is a nontrivial problem. Appendix A presents a general method to determine
whether a sequence is legal for a given $\left(b,\tau\right)$, which
can also be used to calculate a solution analytically from a sequence
for a given $\left(b,\tau\right)$. This method was used to plot the
solutions in Figs.~\ref{fig:symbolic},\ref{fig:sn_bif} and the
bifurcation curves in Fig.~\ref{fig:bifurcations}.

\subsection{Torus bifurcation of the $1\mathbin{:}1$ solution}

We now apply our sequence representation in analysing the $1\mathbin{:}1$
solution; specifically, we determine what happens at the vertical
black lines along the boundary of the locked region in Fig.~\ref{fig:bifurcations}.
Consider the set of half-sequences that represent the $1\mathbin{:}1$
solution. Each half-sequence contains one $Z$ or $\bar{Z}$, one
$H$ or $\bar{H}$, and one $D$ or $\bar{D}$. We need only consider
half-sequences starting with $Z$, as $\left[X_{1},X_{2},X_{3}\right]^{-}=\left[X_{3},X_{1},X_{2}\right]^{-}=\left[X_{2},X_{3},X_{1}\right]^{-}$.
There are only eight possibilities: $\left[Z,H,D\right]^{-}$, $\left[Z,D,H\right]^{-}$,
$\left[Z,\bar{H},D\right]^{-}$, $\left[Z,D,\bar{H}\right]^{-}$,
$\left[Z,H,\bar{D}\right]^{-}$, $\left[Z,\bar{D},H\right]^{-}$,
$\left[Z,\bar{H},\bar{D}\right]^{-}$, $\left[Z,\bar{D},\bar{H}\right]^{-}$.
$\left[Z,H,D\right]^{-}$and $\left[Z,D,H\right]^{-}$ are not legal,
as they require $Z$ to occur when $\dot{x}<0$, which is impossible.
Similarly, $\left[Z,\bar{H},D\right]^{-}$and $\left[Z,D,\bar{H}\right]^{-}$are
legal only for $b<1$, and $\left[Z,\bar{D},H\right]^{-}$and $\left[Z,H,\bar{D}\right]^{-}$are
legal only for $b>1$. The $1\mathbin{:}1$ solutions shown in Fig.~\ref{fig:sample_solutions}
are $\left[Z,\bar{H},\bar{D}\right]^{-}$ in Fig.~\ref{fig:sample_solutions}(d)
and $\left[Z,\bar{D},\bar{H}\right]^{-}$ in Fig.~\ref{fig:sample_solutions}(g).
In the case $b>1$, the sequence representing the $1\mathbin{:}1$
solution is restricted by $\tau$ in the following way:

\begin{equation}
\begin{aligned}\left[Z,\bar{H},\bar{D},\bar{Z},H,D\right]\text{ for } & \tau\bmod1\in\left[0,0.25\right),\\
\left[Z,\bar{D},\bar{H},\bar{Z},D,H\right]\text{ for } & \tau\bmod1\in\left[0.25,0.25\right),\\
\left[Z,H,\bar{D},\bar{Z},\bar{H},D\right]\text{ for } & \tau\bmod1\in\left[0.5,0.25\right),\\
\left[Z,\bar{D},H,\bar{Z},D,\bar{H}\right]\text{ for } & \ensuremath{\tau\bmod1\in\left[0.75,1\right)}.
\end{aligned}
\label{eq:7}
\end{equation}
This can be explained by observing that for small $\tau$, $\bar{H}$
must follow the $Z$ that created it almost immediately. As $\tau$
increases, $\bar{H}$ drifts further away from $Z$ in the sequence,
drifting past $\bar{D}$ at $\tau=0.25$. At $\tau=0.5$, $\bar{H}$
drifts past the subsequent $\bar{Z}$. At $\tau=0.75$, $\bar{H}$
drifts past the subsequent $D$. At $\tau=1$, the $\bar{H}$ created
at $Z$ drifts past the subsequent $Z$, and the pattern repeats.
At the same time, the same drift pattern occurs between $\bar{Z}$
and $H$. We now apply this information to analyse the $1\mathbin{:}1$
solution.

We construct a Poincaré map on the $1\mathbin{:}1$ solution. Let
$t_{z}$ be a time on the $1\mathbin{:}1$ solution at which $x=0$;
then the state of the system at time $t_{z}$ is
\begin{equation}
S(t_{z})=\left(0;z_{0},z_{1},...,z_{n-2},t_{z}\right)^{T}.
\end{equation}
As $x=0$ at $t=t_{z}$, we drop $x$ and rewrite the state as 
\begin{equation}
S_{z}=\left(z_{0},z_{1},...,z_{n-2},t_{z}\right)^{T}.
\end{equation}
Let $t_{z}^{*}$ be the time on the $1\mathbin{:}1$ solution at which
$x=0$ immediately after $t_{z}$. Then the state of the system at
time $t_{z}^{*}$ is
\begin{equation}
S_{z}^{*}=\left(z_{1},z_{2},...,t_{z},t_{z}^{*}\right)^{T}.
\end{equation}
As the $1\mathbin{:}1$ solution is invariant under the symmetry $x\left(t\right)\rightarrow-x\left(t+\frac{1}{2}\right)$,
it has the property
\begin{equation}
S_{z}=\left(\begin{array}{c}
z_{0}\\
z_{1}\\
...\\
z_{n-2}\\
t_{z}
\end{array}\right)=\left(\begin{array}{c}
z_{1}-\frac{1}{2}\\
z_{2}-\frac{1}{2}\\
...\\
t_{z}-\frac{1}{2}\\
t_{z}^{*}-\frac{1}{2}
\end{array}\right)=S_{z}^{*}-\frac{1}{2}.\label{eq:11}
\end{equation}
We define a Poincaré map $\mathbb{P}:S_{z}\rightarrow S_{z}^{*}-\frac{1}{2}$
so that the $1\mathbin{:}1$ solution is a fixed point of $\mathbb{P}$.
Therefore, $\mathbb{P}$ is defined by
\begin{equation}
\mathbb{P}\left(\begin{array}{c}
z_{0}\\
z_{1}\\
...\\
z_{n-2}\\
t_{z}
\end{array}\right)=\left(\begin{array}{c}
z_{1}-\frac{1}{2}\\
z_{2}-\frac{1}{2}\\
...\\
t_{z}-\frac{1}{2}\\
t_{z}^{*}-\frac{1}{2}
\end{array}\right).\label{eq:14}
\end{equation}
For the purpose of deriving $\mathbb{P}$, we will assume w.l.o.g.
that the trajectory is transitioning from $x<0$ to $x>0$ at $t_{z}\in\left[0,0.5\right)$.
Let $t_{h}\in\left[t_{z},t_{z}^{*}\right)$ be the time at which the
feedback changes. First, we consider the case where $\tau<0.5$; then
the zero element generated at $t_{z}$ is consumed at $t_{h}=t_{z}+\tau$.
Then $\mathbb{P}$ is one-dimensional as $S_{z}$ has only one variable,
$t_{z}$. We write the one-dimensional map as
\begin{equation}
\mathbb{P}\left(t_{z}\right)=t_{z}^{*}-\frac{1}{2}.\label{eq:13-1}
\end{equation}
If $\tau<0.5$, then the solution is represented by the sequence $\left[Z,\bar{H},\bar{D}\right]^{-}$
; therefore, the feedback is positive for $t\in[t_{z},t_{h})$. We
write an equation for the displacement of the trajectory between $t_{z}$
and $t_{z}^{*}$ as
\begin{equation}
b\left(\frac{1}{2}-t_{z}\right)-b\left(t_{z}^{*}-\frac{1}{2}\right)+(t_{h}-t_{z})-\left(t_{z}^{*}-t_{h}\right)=0.\label{eq:9}
\end{equation}
We substitute $t_{h}=t_{z}+\tau$ and solve for $t_{z}^{*}-\frac{1}{2}$
to obtain
\begin{equation}
\mathbb{P}(t_{z})=-\left(\frac{b-1}{b+1}\right)t_{z}+\frac{b+2\tau}{b+1}-\frac{1}{2}.
\end{equation}
As $\left|\frac{b-1}{b+1}\right|<1$ for $b\geq0$, $\mathbb{P}$
is stable for $\tau<\frac{1}{2}$. 

If $\tau\in\left[0.5,1\right)$, then $t_{h}$ was generated, not
at $t_{z}$, but at the previous $x=0$ crossing $z_{0}$. Then $\mathbb{P}$
is two-dimensional as $S_{z}$ has two variables, and the feedback
is negative for $t\in\left[t_{z},t_{h}\right)$, where $t_{h}=z_{0}+\tau$.
These conditions can be generalised for larger $\tau$. The dimension
of the system is $n=\left\lceil 2\tau\right\rceil $, where $\left\lceil 2\tau\right\rceil $
is the smallest integer greater than $2\tau$; then the feedback for
$t\in\left[t_{z},t_{h}\right)$ is $(-1)^{n-1}$. We can then generalise
Eq.~(\ref{eq:9}) for arbitrary $n$ and solve for $t_{z}^{*}-\frac{1}{2}$
to obtain
\begin{equation}
t_{z}^{*}-\frac{1}{2}=-t_{z}+\frac{b+2t_{h}(-1)^{n-1}}{b+(-1)^{n-1}}-\frac{1}{2}.\label{eq:13}
\end{equation}
For $\tau>\frac{1}{2}$, $\mathbb{P}$ can written as 
\begin{equation}
\mathbb{P}\left(S_{z}\right)=AS_{z}+B,
\end{equation}
where $A$ is an $n\times n$ matrix
\begin{equation}
A=\left(\begin{array}{ccccc}
0 & 1 & 0 & \cdots & 0\\
\vdots & 0 & \ddots & \ddots & \vdots\\
\vdots & \vdots & \ddots & \ddots & 0\\
0 & 0 & \cdots & 0 & 1\\
\frac{2(-1)^{n-1}}{b+(-1)^{n-1}} & 0 & \cdots & \cdots & -1
\end{array}\right),
\end{equation}
$B$ is given in Appendix B. Note that $A$ only depends on the parameter
$b$ and not on $\tau$. This means that for fixed $n$, the value
of of $b$ at which the fixed point of $\mathbb{P}$ is bifurcating,
$b_{\text{bif}}$, is constant. By solving the characteristic equation,
we calculate 
\begin{equation}
b_{\text{bif}}(n)=\frac{1}{\cos\left(\frac{\pi\left(n-1\right)}{2n-1}\right)}-(-1)^{n-1},
\end{equation}
where $n=\left\lceil 2\tau\right\rceil $. Full calculations may be
found in Appendix B. The curve $\left(b_{\text{bif}}\left(\left\lceil 2\tau\right\rceil \right),\tau\right)$
can be seen plotted as black vertical lines against maximum charts
in Fig.~\ref{fig:bifurcations}\emph{.} If we examine the maximum
chart where $b$ is swept from right to left, we see that the system
ceases to converge to the $1\mathbin{:}1$ solution along this curve.
We now know that this is because the fixed point of $\mathbb{P}$,
and hence the $1\mathbin{:}1$ solution, loses stability at $b=b_{\text{bif}}$.
By calculating the eigenvalues of $\mathbb{P}$ explicitly for $n=2$
and $n=3$, we find that the loss of stability occurs because a pair
of complex conjugate eigenvalues $\lambda_{1,2}$ cross $|\lambda|=1$.
Therefore the fixed point of $\mathbb{P}$ loses stability due to
a Neimark-Sacker (NS) bifurcation\citep{hone_neimarksacker_2010}.
As $\mathbb{P}$ is a Poincaré map on a periodic orbit, a NS bifurcation
of the fixed point of $\mathbb{P}$ corresponds to a torus (T) bifurcation
of the $1\mathbin{:}1$ solution. However, $\mathbb{P}$ only shows
the existence of the T bifurcation along the vertical sections of
the boundary of the locked region, where $n$ is constant and $\mathbb{P}$
is smooth. To fully understand the horizontal sections of the boundary,
we must look to non-smooth bifurcation theory.

\section{Border-collision bifurcations}

The Arnold tongues seen in Fig.~\ref{fig:period} have sharply defined
boundaries, made up of curves and straight lines. We seek to determine
what happens to solutions at these boundaries and derive analytic
expressions for the boundaries using Poincaré maps.

By simulating solutions near the boundaries of the Arnold tongues,
we observe that moving closer to the boundaries causes a $D$ or $\bar{D}$
to move closer to $x=0$. An example of this is Fig.~\ref{fig:symbolic}(a),
where the $D$ at $t=0$ in the $5\mathbin{:}1$ solution occurs for
$x>0$ near $x=0$. If this $D$ crossed $x=0$ and occurred at $x<0$,
this would significantly impact the feedback. There would be two additional
$x=0$ crossings in the trajectory, changing the $D$ to $\bar{Z}DZ$
and adding a $H$ and a $\bar{H}$ elsewhere in the sequence. Therefore,
when we construct a Poincaré map to describe the dynamics of the system
close to the boundary of such a tongue, the map must have a border
at $x=0$, such that the map is continuous across the border but not
differentiable at the border. Such maps, and the associated border-collision
bifurcations, have recently received systematic analysis in the literature\citep{bernardo_bifurcations_2002,banerjee_border_1999,colombo_bifurcations_2012,di_bernardo_bifurcations_2008,granados_border_2014,meiss_neimarksacker_2008,nusse_border-collision_1994,holmberg_relay_1993,colombo_complex_2007}.

\subsection{Border-collision saddle-node bifurcation of the $5\mathbin{:}1$
and $5\mathbin{:}3$ solutions}

\begin{figure}[!t]
\includegraphics[width=1\columnwidth]{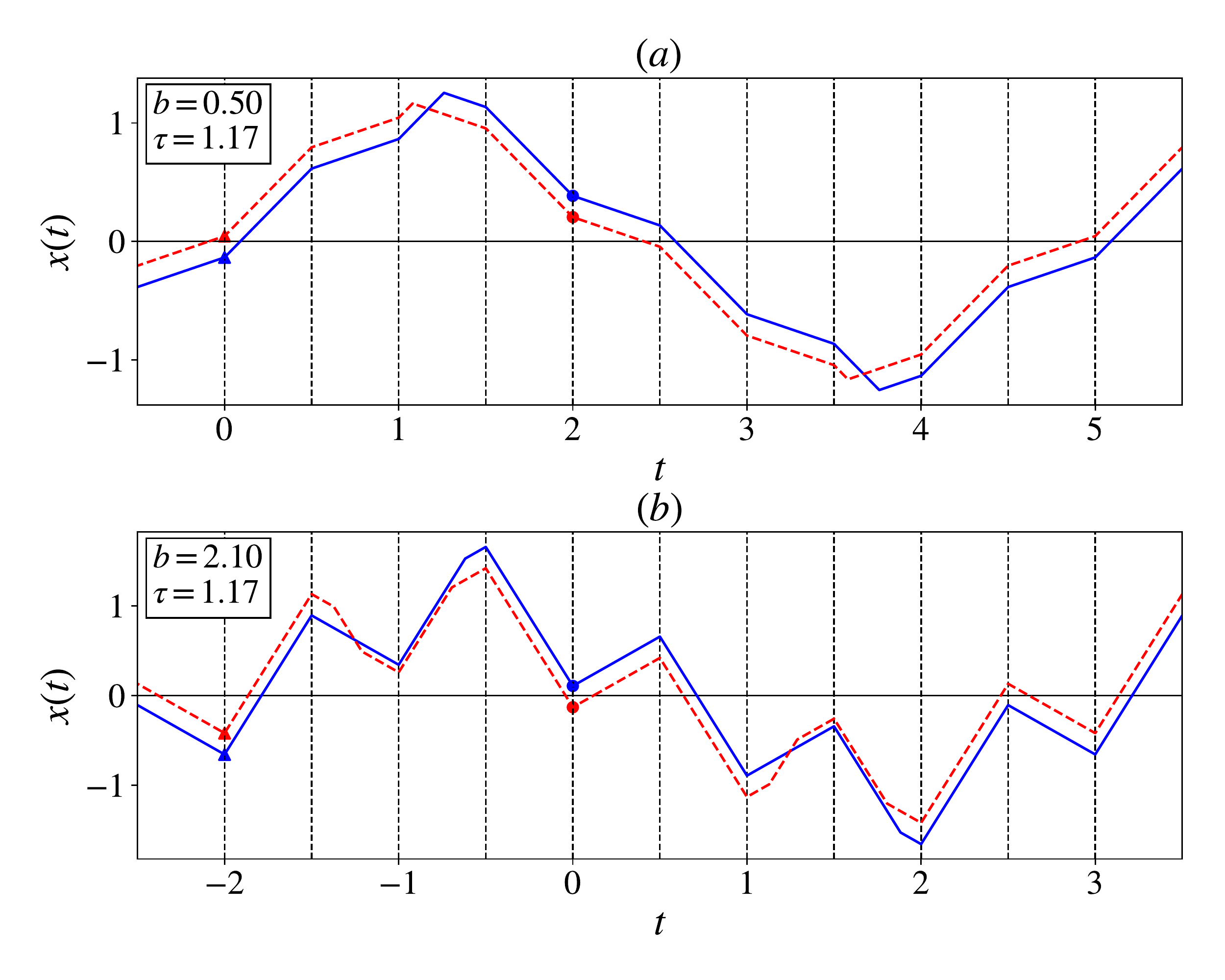}

\includegraphics[width=1\columnwidth]{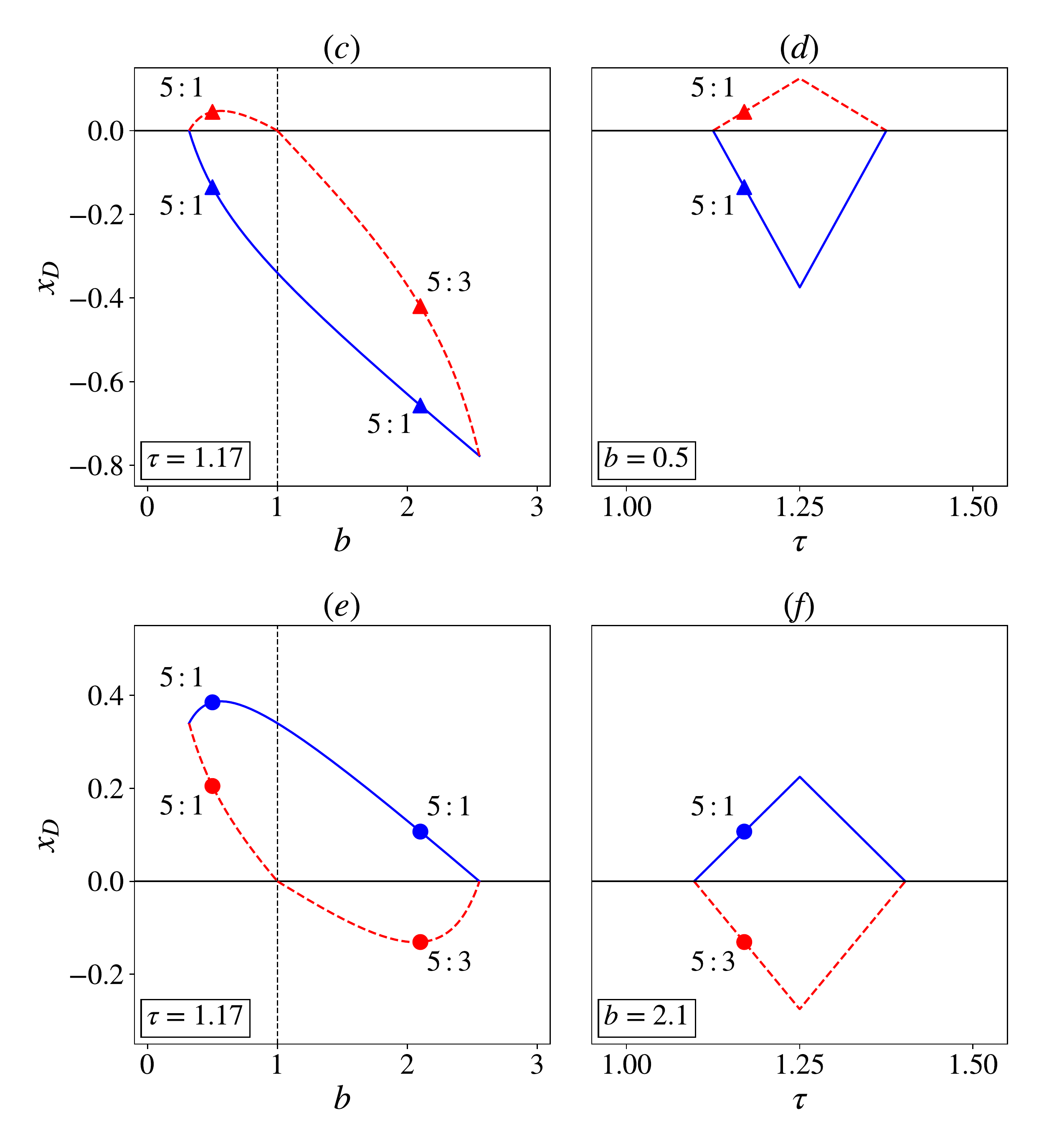}

\caption{\label{fig:sn_bif} BCSN bifurcations at the boundary of the $5\mathbin{:}1$
tongue. The stable solution is plotted in solid blue, while the unstable
solution is plotted in dashed red. In $(a)$, the stable $5\mathbin{:}1$
solution $\left[Z,\bar{D},D,\bar{H},\bar{D},D,\bar{D}\right]^{-}$
and the unstable $5\mathbin{:}1$ solution $\left[Z,D,\bar{D},D,\bar{H},\bar{D},D\right]^{-}$
are plotted close to $\left(b,\frac{5-b}{4}\right)$. In $(b)$, the
stable $5\mathbin{:}1$ solution $\left[Z,\bar{D},D,\bar{H},\bar{D},D,\bar{D}\right]^{-}$
and the unstable $5\mathbin{:}3$ solution $\left[Z,\bar{D},\bar{H},H,D,\bar{H},\bar{D},\bar{Z},D,Z,\bar{D}\right]^{-}$
are plotted close to $\left(b,\frac{b^{2}+2b+5}{4\left(b+1\right)}\right)$.
The $D$ events associated with the BCSN bifurcation for $b<1$ are
plotted as triangles, while those associated with the BCSN bifurcation
for $b>1$ are plotted as circles; $(c)$, $(d)$, $(e)$ and $(f)$
show their evolution as $b$ and $\tau$ vary.}

\end{figure}

We construct a Poincaré map $\mathbb{B}$ on the $5\mathbin{:}1$
and $5\mathbin{:}3$ solutions represented by $\left[Z,\bar{D},D,\bar{H},\bar{D},D,\bar{D}\right]^{-}$
and $\left[Z,\bar{D},\bar{H},H,D,\bar{H},\bar{D},\bar{Z},D,Z,\bar{D}\right]^{-}$
respectively, which can be seen in Fig.~\ref{fig:sn_bif}(a), such
that our fixed point is the state of the system at time $t=0$, at
the position $x_{D}$ at which the $D$ that will cross $x=0$ occurs.
As the $\bar{H}$ created at $Z$ occurs before this $D$, $\mathbb{B}$
is one-dimensional. We divide $\mathbb{B}$ into $\mathbb{B}^{+}$
and $\mathbb{B}^{-}$ for $x\geq0$ and $x<0$ respectively. The $5\mathbin{:}1$
and $5\mathbin{:}3$ solutions are invariant under the symmetry $x(t)=-x\left(t+\frac{1}{2}\right)$;
therefore we construct our map $\mathbb{B}:x_{D}\left(0\right)\rightarrow-x_{D}\left(\frac{5}{2}\right)$.
$\mathbb{B^{\mathrm{+}}}$ is a map on the solution represented by
$\left[Z,\bar{D},D,\bar{H},\bar{D},D,\bar{D}\right]^{-}$. Following
the blue curve in Fig.~\ref{fig:sn_bif}(b), we can derive
\begin{equation}
\mathbb{B}^{+}(x_{D})=-\left(\frac{b-1}{b+1}\right)x_{D}+\frac{2b}{b+1}-\frac{b+5}{2}+2\tau.
\end{equation}
$\mathbb{B}^{-}$ is a map on the $5\mathbin{:}3$ solution represented
by $\left[Z,\bar{D},\bar{H},H,D,\bar{H},\bar{D},\bar{Z},D,Z,\bar{D}\right]^{-}$,
which is shown in Fig.~\ref{fig:sn_bif}(b) in red. The feedback
change due the trajectory dipping below $x=0$ occurs at some $t\in\left(1,1.5\right)$
has duration $\frac{-2bx}{b^{2}-1}$; $\mathbb{B}^{-}$ is otherwise
identical to $\mathbb{B^{\mathrm{+}}}$. Thus we derive $\mathbb{B}$
as
\begin{equation}
\mathbb{B}(x_{D})=\begin{cases}
-\left(\frac{b-1}{b+1}\right)x_{D}+\frac{2b}{b+1}-\frac{b+5}{2}+2\tau & x_{D}\geq0\\
\left(\frac{4b}{b^{2}-1}-\frac{b-1}{b+1}\right)x_{D}+\frac{2b}{b+1}-\frac{b+5}{2}+2\tau & x_{D}<0
\end{cases}
\end{equation}
By setting $x_{D}=0$ and solving $\mathbb{B}$ for $\tau$, we obtain
the curve $\tau=\frac{b^{2}+2b+5}{4\left(b+1\right)}$. This matches
the lower right boundary of the $5\mathbin{:}1$  tongue spanning
from $\left(1,1\right)$ to $\left(3,1.25\right)$. For $b\in\left[1,3\right]$
and $\tau\in\left[\frac{b^{2}+2b+5}{4\left(b+1\right)},1.25\right]$,
$\mathbb{B}$ has two fixed points; a stable fixed point that exists
for $x_{D}\geq0$ and an unstable fixed point that exists for $x_{D}\leq0$.
These two fixed points collide and vanish at the border $x_{D}=0$
at $\tau=\frac{b^{2}+2b+5}{4\left(b+1\right)}$ in a border-collision
saddle-node (BCSN) bifurcation \citep{banerjee_border_1999}. Hence
the $5\mathbin{:}1$ tongue is bounded by a BCSN bifurcation. Fig.~\ref{fig:sn_bif}(b,e,f)
shows the BCSN bifurcation of the $5\mathbin{:}1$ and $5\mathbin{:}3$
solutions.

We generalise $\mathbb{B}$ for every $P\mathbin{:}1$  tongue for
odd $P\geq5$ as 
\begin{equation}
\mathbb{B}_{P}(x_{D})=\begin{cases}
-\left(\frac{b-1}{b+1}\right)x_{D}+\frac{2b}{b+1}-\frac{b+\left|P-4\tau\right|}{2} & x_{D}\geq0\\
\left(\frac{4b}{b^{2}-1}-\frac{b-1}{b+1}\right)x_{D}+\frac{2b}{b+1}-\frac{b+\left|P-4\tau\right|}{2} & x_{D}<0
\end{cases}
\end{equation}
where $\left|P-4\tau\right|$ is the term that causes the $P\mathbin{:}1$
tongues to be symmetric across $\tau=\frac{P}{4}$. By setting $x_{D}=0$,
we calculate the boundaries of all $P\mathbin{:}1$  tongue for odd
$P\geq5$ and $b\in\left[1,3\right]$ as
\begin{equation}
\mathbb{\tau_{\mathrm{P}}}(b)=\frac{P}{4}\pm\frac{3b-b^{2}}{4\left(b+1\right)}.
\end{equation}
We find that this phenomenon persists throughout the system. The vast
majority of tongues investigated are bounded by BCSN bifurcations
occurring when a $D$ crosses $x=0$. For $b>1$, a stable $P\mathbin{:}R$
solution and an unstable $P\mathbin{:}R+2$ solution undergo a SN
bifurcation when a $D$ crosses $x=0$ if the solution has the symmetry
$x(t)=-x(t+\frac{P}{2})$. Otherwise, the stable $P\mathbin{:}R$
solution undergoes a SN bifurcation with an unstable $P\mathbin{:}R+1$
solution when a $D$ crosses $x=0$. 

For $b<1$, the mechanism by which BCSN bifurcations occur is slightly
different. Instead of a $D$ crossing $x=0$ and adding new symbols
to the sequence, $D$ instead crosses $x=0$ by swapping order with
an existing $Z$. The stable $5\mathbin{:}1$ solution $\left[Z,\bar{D},D,\bar{H},\bar{D},D,\bar{D}\right]^{-}$undergoes
a BCSN bifurcation with the unstable $5\mathbin{:}1$ $\left[Z,D,\bar{D},D,\bar{H},\bar{D},D\right]^{-}$at
$\left(b,\frac{5-b}{4}\right)$, $b<1$, as shown in Fig.~\ref{fig:sn_bif}(a,c,d).
While the unstable $5:1$ and $5:3$ solutions shown in Fig. 5 are
two different solutions under our sequence representation, they are
identical at $b=1$, where a $D$ passes through $x=0$ without a
bifurcation. A noteworthy feature of the BCSN bifurcations at the
boundary of the $5\mathbin{:}1$ tongue is that the pair of $D$ events
which collide at $x=0$ is different for $b<1$ and $b>1$, as seen
in Figure Fig.~\ref{fig:sn_bif}(a,b,c,e). Generally, for $b<1$,
a stable $P\mathbin{:}1$ solution undergoes a BCSN bifurcation with
an unstable $P\mathbin{:}1$ solution. Again, we produce a general
form for the curves where a $P\mathbin{:}1$ solution undergoes a
BCSN bifurcation for odd $P\geq1$ and and $b\in\left[0,1\right]$
as 
\begin{equation}
\mathbb{\tau_{\mathrm{P}}}(b)=\frac{P\pm b}{4}.
\end{equation}
 The $1\mathbin{:}1$ solution is a special case. The stable $\left[Z,\bar{H},\bar{D}\right]^{-}$
solution and the unstable $\left[Z,D,\bar{H}\right]^{-}$ undergo
a BCSN bifurcation along $\left(b,\frac{n}{2}+\frac{1-b}{4}\right)$,
$b<1$, $n\in\mathbb{Z^{\mathrm{+}}}$ . The stable $\left[Z,\bar{D},\bar{H}\right]^{-}$
solution and the unstable $\left[Z,\bar{H},D\right]^{-}$ undergo
a BCSN bifurcation along $\left(b,\frac{n}{2}+\frac{1+b}{4}\right)$,
$b<1$, $n\in\mathbb{Z^{\mathrm{+}}}$.

A selection of these curves can be seen plotted in solid white in
Fig.~\ref{fig:bifurcations}. There is a conspicuous outlier to this
pattern. The $3\mathbin{:}1$ solution does not undergo a BCSN bifurcation
at the right boundary of the $3\mathbin{:}1$  tongue for $b>1$.
We will now investigate why this is the case.

\subsection{Border-collision torus bifurcation of the $3\mathbin{:}1$ solution}

Fig.~\ref{fig:period} shows that the $3\mathbin{:}1$  tongue has
the same shape as the other $P\mathbin{:}1$  tongues, rooted at $\tau=0.75$.
Along some of the boundary, the $3\mathbin{:}1$ solution changes
to the $3\mathbin{:}3$ solution without bifurcation. We test the
observed sequences which represent the $3\mathbin{:}3$ solution and
find that the $3\mathbin{:}3$ solution exists in the region $\left[1,3\right]\times\left[0.5,1\right]$
outside of the $3\mathbin{:}1$  tongue. However, the $3\mathbin{:}3$
solution is not always stable; this can be seen in the top half of
the $3\mathbin{:}3$ region in Fig.~\ref{fig:period}. In order to
investigate this phenomenon, we construct a Poincaré map $\mathbb{T}$
by a similar method as for the BCSN bifurcation. For $\tau>0.75$,
the $3\mathbin{:}1$ solution is represented by $\left[Z,\bar{D},D,\bar{H},\bar{D}\right]^{-}$
from which we derive $\mathbb{T}^{+}$ for $x_{D}\geq0$. For $x_{D}<0$,
the sequence changes to $\left[Z,\bar{H},H,\bar{D},\bar{Z},D,Z,\bar{H},\bar{D}\right]^{-}$,
from which we derive $\mathbb{T}^{-}$. There is a significant difference
however; the $3\mathbin{:}1$ solution is one-dimensional, but for
$\tau>0.75$, the $3\mathbin{:}3$ solution is two-dimensional. This
occurs because the additional $\bar{H}$ and $H$ events occur between
the $D$ that crosses $x=0$ and the $Z$ present in both sequences.
Therefore, we also need to know the time $t_{\bar{H}}$ at which the
$\bar{H}$ present in both sequences occurs. We derive $\mathbb{T}:\left(x_{D},t_{\bar{H}}\right)^{T}\rightarrow\left(x_{D}^{*},t_{\bar{H}}^{*}-\frac{3}{2}\right)^{T}$
as
\begin{equation}
\mathbb{T}\left(\begin{array}{c}
x_{D}\\
t_{\bar{H}}
\end{array}\right)=\begin{cases}
\left(\begin{array}{cc}
-1 & -2\\
\frac{1}{b+1} & \frac{2}{b+1}
\end{array}\right)\left(\begin{array}{c}
x_{D}\\
t_{\bar{H}}
\end{array}\right)+C & x_{D}\geq0\\
\left(\begin{array}{cc}
\frac{4b}{b^{2}-1}-1 & -2\\
\frac{1}{b+1} & \frac{2}{b+1}
\end{array}\right)\left(\begin{array}{c}
x_{D}\\
t_{\bar{H}}
\end{array}\right)+C & x_{D}<0
\end{cases}
\end{equation}
where $C=\left(\begin{array}{c}
\frac{3-b}{2}\\
\frac{b}{b+1}+\tau-\frac{3}{2}
\end{array}\right)$. Note that $\mathbb{T}^{+}$ has two rows which are multiples of
each other; this is due to the $3\mathbin{:}1$ solution being one-dimensional,
and results in a $0$ eigenvalue. By setting $x_{D}=0$ and solving
$\mathbb{T}$ for $\tau$, we obtain the curve $\tau=\frac{3}{4}+\frac{3b-b^{2}}{4\left(b+1\right)}$,
which matches the upper right  boundary of the $3\mathbin{:}1$  tongue
spanning from $\left(1,1\right)$ to $\left(3,0.75\right)$. $\mathbb{T}$
has a single fixed point which crosses $x_{D}=0$ at the boundary
of the $3\mathbin{:}1$ tongue. For $x_{D}>0$ the fixed point is
always stable. For $x_{D}<0$ the fixed point is stable for $b>2.6038$,
and loses stability when a pair of complex conjugate eigenvalues pass
through $\left|\lambda\right|=1$, resulting in a NS bifurcation at
$b=2.6038$. $\left|\lambda_{\pm}\right|<1$ for $b>2.6038$. Therefore,
for $\tau>0.75$ the $3\mathbin{:}3$ solution is stable for $b>2.6038$.
This agrees with the shape of the region in which the $3\mathbin{:}3$
solution is observed in Fig.~\ref{fig:period}. 

\begin{figure}[!t]
\includegraphics[width=1\columnwidth]{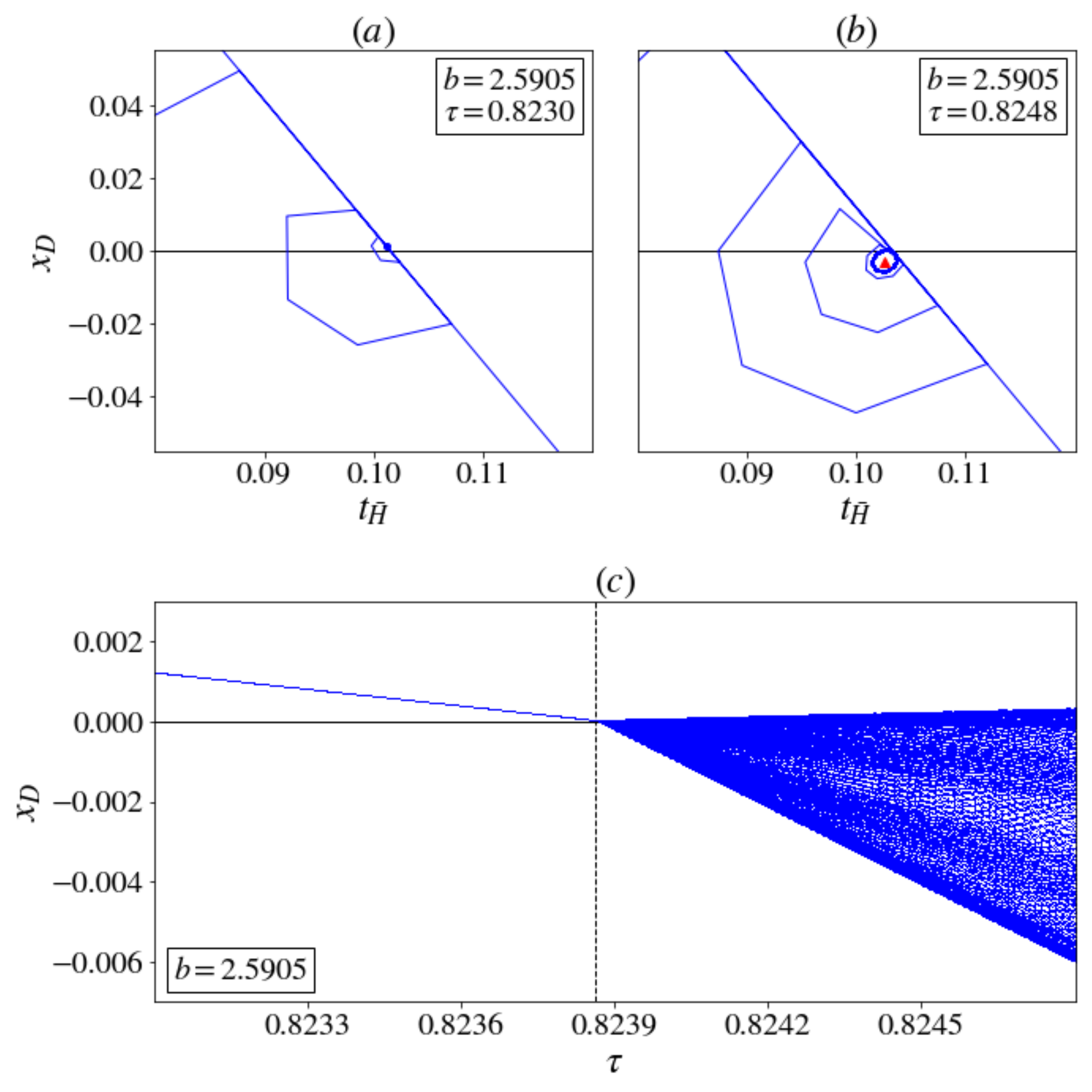}

\caption{\label{fig:bct} Border-collision Neimark-Sacker (BCNS) bifurcation
of the fixed point of the $3\mathbin{:}1$ solution. The upper two
plots show the BCNS bifurcation of the half-period map $\mathbb{T}$.
The lower plot is based on simulation using our iterative map, and
shows the stable closed invariant curve expanding from the fixed point
from $x=0$. The fixed point is plotted as a blue circle where it
is stable, and as a red triangle where it is unstable.}
\end{figure}
 For $b<2.6038$, we observe an interesting interaction between $\mathbb{T}^{+}$
and $\mathbb{T}^{-}$. As $\mathbb{T}^{+}$ has a $0$ eigenvalue,
any point $\left(x_{D},t_{\bar{H}}\right)$ for $x_{D}>0$ is mapped
directly to a nullcline on which the fixed point sits for $x_{D}>0$.
The trajectory then undergoes decaying oscillations to the fixed point
in the nullcline. However, if the fixed point is close to $x_{D}=0$,
$\mathbb{T}^{+}$ can map the point into $x_{D}<0$, where the nullcline
does not exist. In the absence of a stable fixed point at $x_{D}<0$,
the trajectory starts to spiral out to infinity. As this spiral must
cross $x_{D}=0$ eventually, the trajectory is caught by the nullcline
again, causing an interesting half-spiral attraction. When the fixed
point is at $x_{D}<0$, the trajectory converges to a stable attractor
that strikes the nullcline at multiple points. Fig.~\ref{fig:bct}(a,b)
show the interaction between $\mathbb{T}^{+}$ and $\mathbb{T}^{-}$.
Fig.~\ref{fig:bct}(c) shows the results of simulations in a $\left(b,\tau\right)$
sweep that crosses the boundary of the $3\mathbin{:}1$ Arnold tongue.
The simulated results agree with those derived from $\mathbb{T}$,
and confirm that the closed invariant curve is born from the fixed
point as it crosses $x_{D}=0$. We consider this to be a border-collision
Neimark-Sacker (BCNS) bifurcation of $\mathbb{T}$, similar to that
seen by \citet{meiss_neimarksacker_2008}. It corresponds to a border-collision
torus (BCT) bifurcation of the $3\mathbin{:}1$ solution in the continuous
system. The BCT bifurcation is supercritical, as a stable closed invariant
curve expands from the fixed point when it becomes unstable. The BCT
bifurcation of the $3\mathbin{:}1$ solution and the T bifurcation
of the $3\mathbin{:}3$ solution are plotted in black in Fig.~\ref{fig:bifurcations}.

\subsection{Border-collision torus bifurcation of the $1\mathbin{:}1$ solution}

\begin{figure}[!t]
\includegraphics[width=1\columnwidth]{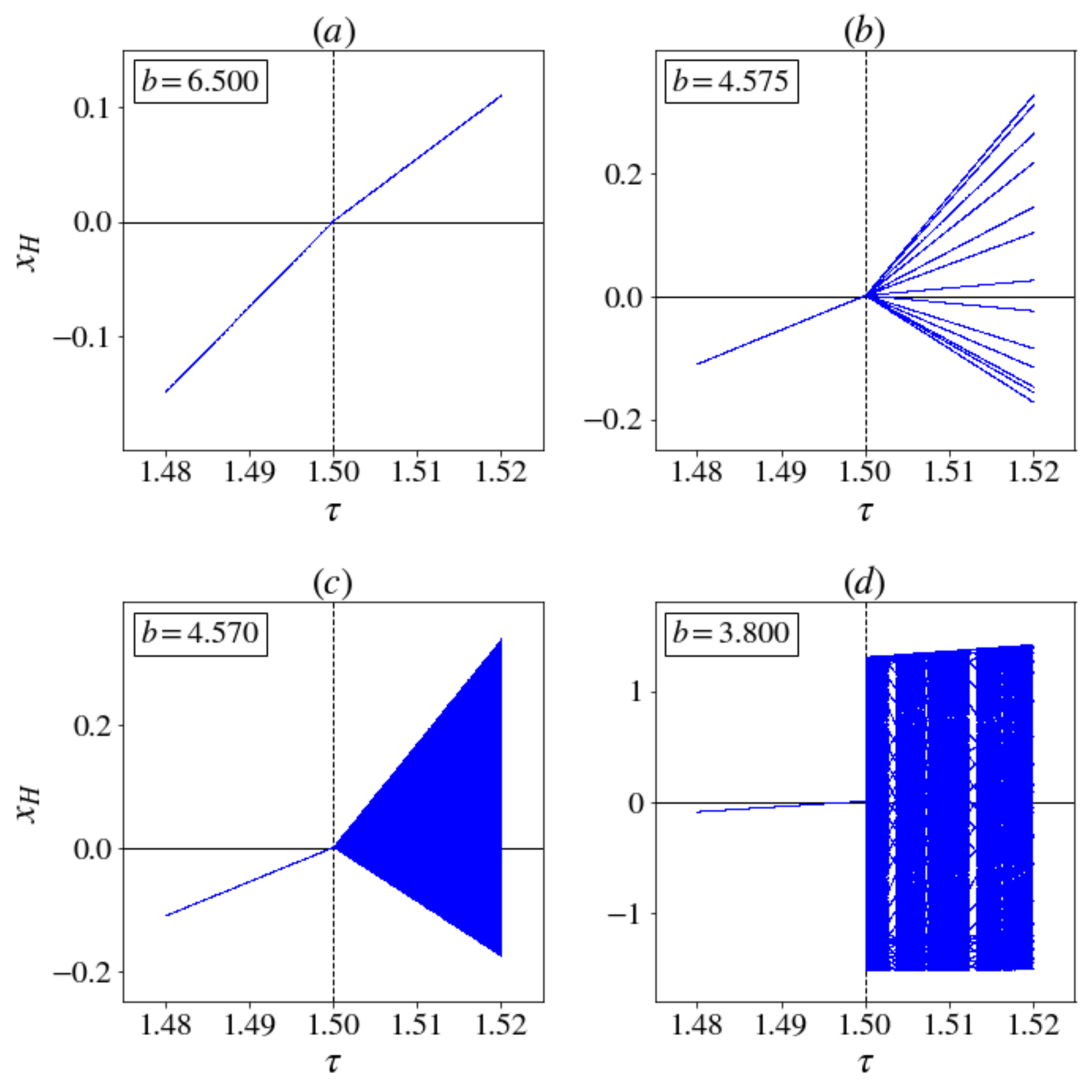}

\caption{\label{fig:bct_locked} x-coordinates of $H$ events from simulations.
The four plots show the different cases of the $1\mathbin{:}1$ solution
at the border.}
\end{figure}

Now that we have studied the BCT bifurcation of the $3\mathbin{:}1$
solution, we return to the $1\mathbin{:}1$ solution. Previously we
noted that the Poincaré map $\mathbb{P}$ only proved the existence
of the torus bifurcation along the vertical boundaries of the locked
region. We can now consider the horizontal boundaries in terms of
border collisions. Recall that the $1\mathbin{:}1$ solution is represented
by four different sequences. At $\tau=\frac{n}{2}$, $n\in\mathbb{N}$,
the sequence representing the $1\mathbin{:}1$ solution changes when
a $H$ and a $\bar{H}$ cross $x=0$. This results in a change in
the dimension of $\mathbb{P}$. Let $\mathbb{P}_{n}$ denote $\mathbb{P}$
when $\mathbb{P}$ contains an $n\times n$ matrix. When $\tau$ increases
past $\tau=\frac{n}{2}$, $\mathbb{P}$ changes from $\mathbb{P}_{n}$
to $\mathbb{P_{\mathrm{n+1}}}$. Let us consider $x_{H}$, the position
of the $H$ event, as the fixed point of the map. Then $x_{H}=0$
when $\mathbb{P}$ changes from $\mathbb{P}_{n}$ to $\mathbb{P_{\mathrm{n+1}}}$,
resulting in a border collision at $x_{H}=0$. Fig.~\ref{fig:bct_locked}
shows four cases that occur in the border collision when we sweep
across the border $\tau=1.5$ for fixed $b$. 

Fig.~\ref{fig:bct_locked}(a) shows the stable $1\mathbin{:}1$ solution
remaining stable, as both $\mathbb{P_{\mathrm{3}}}$ and $\mathbb{P_{\mathrm{4}}}$
are stable at $b=6.5$. Fig.~\ref{fig:bct_locked}(b) shows the stable
$1\mathbin{:}1$ solution becoming unstable, generating a $13\mathbin{:}13$
solution in a supercritical BCT bifurcation. This solution exists
in one of the vertical stripes we noted in Fig.~\ref{fig:period}.
Fig.~\ref{fig:bct_locked}(c) shows the stable $1\mathbin{:}1$ solution
becoming unstable, generating an aperiodic solution in a supercritical
BCT bifurcation, showing that there are gaps between the vertical
stripes. Fig.~\ref{fig:bct_locked}(d) shows the stable $1\mathbin{:}1$
solution becoming unstable; however, the aperiodic solution that the
system converges to afterwards is not generated at the border, indicating
that the BCT bifurcation is subcritical at this point. Therefore the
BCT bifurcation of the $1\mathbin{:}1$ solution must change criticality
at some point along $\tau=1.5$. The BCT bifurcation of the $1\mathbin{:}1$
solution produces the horizontal black lines in Fig.~\ref{fig:bifurcations},
completing the boundary of the locked region.

We examine the difference in maxima of nearby solutions on either
side of $\tau=1.5$ in Fig.~\ref{fig:bifurcations}(b). The point
$\left(b,\tau\right)=\left(1.5,3.86\right)$ stands out; the difference
in maxima is sudden to the left of that point, and gradual to the
right. The gradual change in maxima occurs where the BCT bifurcation
is supercritical, and the abrupt change in maxima occurs where the
BCT bifurcation is subcritical, as illustrated by Fig.~\ref{fig:bct_locked}(b-d).
We note that $\left(b,\tau\right)=\left(1.5,3.86\right)$ forms one
corner of a roughly triangular region containing vertical $P\mathbin{:}P$
stripes seen in Fig.~\ref{fig:period}; every $P\mathbin{:}R$ solution
in this region is $P\mathbin{:}P$. This leads us to the dashed white
curve in Fig.~\ref{fig:bifurcations}. We refer to this curve as
the $D\bar{D}$ curve. To the right of the $D\bar{D}$ curve, all
$D$ events occur at $x<0$ and all $\bar{D}$ events occur at $x>0$,
and so a legal sequence cannot contain a $D$ adjacent to a $\bar{D}$;
a $P\mathbin{:}R$ solution in this region must, therefore, be $P\mathbin{:}P$.
To the left of the $D\bar{D}$ curve, a legal sequence representing
a solution can contain a $D$ adjacent to a $\bar{D}$; we refer to
such a solution as a $D\bar{D}$ solution. The point at which the
BCT bifurcation changes criticality occurs where the $D\bar{D}$ curve
intersects the boundary of the locked region. 

As noted previously, when sweeping from right to left, the system
converges to the $1\mathbin{:}1$ solution until it becomes unstable
at the T bifurcation. Now, we see that when sweeping from left to
right, the system converges to $D\bar{D}$ solutions until they vanish
at the $D\bar{D}$ curve; we note that a $6\mathbin{:}5$ and a $6\mathbin{:}6$
solution undergo a BCSN bifurcation at the $D\bar{D}$ curve. If the
T bifurcation lies to the left of the $D\bar{D}$ curve, then the
$1\mathbin{:}1$ solution remains stable while the $D\bar{D}$ solutions
exist, resulting in regions of multistability. We observe that this
behaviour, together with the change in criticality of the BCT bifurcation,
is reminiscent of a Chenciner bifurcation, a co-dimension-2 bifurcation
that was observed in the smooth system by \citet{keane_chenciner_2018}
where it produced rich dynamics.

If the T bifurcation lies to the right of the $D\bar{D}$ curve, this
produces regions where vertical stripes can occur between the BCT
bifurcation of the $1\mathbin{:}1$ solution and the $D\bar{D}$ curve.
Note that the vertical stripes do not stretch all the way between
the $D\bar{D}$ curve and the BCT bifurcation. There is a second condition
that a region must satisfy for the existence of vertical stripes,
which is shown by the dashed black curve in Fig.~\ref{fig:bifurcations}.
Vertical stripes only occur where the order of $D$, $\bar{D}$, $H$,
$\bar{H}$ symbols in the stable solution is consistent with the $1\mathbin{:}1$
solution. The difference between stable $P\mathbin{:}P$ solutions
and the unstable $1\mathbin{:}1$ solution they coexist with lies
only in the order of $Z$ $H$, $\bar{Z}$ and $\bar{H}$ symbols.
This arises because these solutions are generated when $Z$, $H$,
$\bar{Z}$ and $\bar{H}$ symbols swapped order in the sequence representation
of the $1\mathbin{:}1$ solution at $\tau=\frac{n}{2}$, $n\in\mathbb{N}$. 

\section{Conclusion}

We thoroughly analysed an elementary two-parameter system which combines
the effects of time-delayed feedback and periodic forcing. In spite
of its simplicity, it demonstrates a complex structure of Arnold tongues
with zero-width shrinking points and a high degree of multistability.
Due to the system being piecewise-linear, we are able to solve the
system analytically using an iterative map. We investigate the existence
and stability of solutions through the development of a symbolic representation
of solutions and the analysis of the subsequently developed Poincaré
and border-collision maps. This analysis reveals that the Arnold tongues
are bounded by curves of border-collision saddle-node bifurcations
of periodic orbits. Additionally, we find curves of torus bifurcations
connected to curves of border-collision torus bifurcations, and investigate
changes in the criticality of these bifurcations. 

Our analysis sheds new light onto previously obtained results in related
smooth systems, particularly the El Niño Southern Oscillation climate
model studied by \citet{keane_delayed_2015}. Comparing the numerically
calculated bifurcation structure found in that paper with the analytically
calculated bifurcation structure found here reveals that the prominent
features of the smooth system (\ref{eq:1},\ref{eq:2}) are preserved
in the non-smooth limit of $\kappa\rightarrow\infty$. Indeed, the
solutions in the smooth system generally appear as ``smoothed out''
counterparts to the piecewise-linear solutions of the non-smooth system.
There are some significant differences. The tongues in the smooth
system are not connected by shrinking points, which is to be expected
in the presence of nonlinearity, according to \citet{simpson_resonance_2010}.
Additionally, the smooth system contains period doubling bifurcations,
which we do not observe in our system, likely due to the loss of nonlinearity
in the limit of $\kappa\rightarrow\infty$. However, the analysis
presented here does provide new insights into dynamics previously
observed numerically\citep{ghil_delay_2008,keane_delayed_2015}. 

Both delayed feedback and periodic forcing are very common mathematical
model ingredients and can be found in a variety of models used to
study, for example, laser dynamics\citep{sorrentino_effects_2015}
and chimera states\citep{semenov_delayed_2016}. The current work
reveals the phenomena which are a genuine consequence of this combination.
Therefore, we expect this work to be of interest to a wide readership.

\section{Acknowledgements}

This research was funded by the Irish Research Council and McAfee
LLC through the Irish Research Council Employment-Based Postgraduate
Programme. We also thank Dr.~Sorcha Healy and the Applied Data Science
team at McAfee for their technical support.

\section{Appendix}

\subsection{Determining legality of a sequence}

Each event in a sequence has a time and a position associated with
it. The times can be used to generate a system of simultaneous equations:
\begin{itemize}
\item Each $D$ occurs at a time $t_{D}=n$, $n\in\mathbb{Z}$
\item Each $\bar{D}$ occurs at a time $t_{\bar{D}}=n+\frac{1}{2}$, $n\in\mathbb{Z}$
\item Each $H$ occurs at a time $t_{H}=t_{\bar{Z}}+\tau$, where $t_{\bar{Z}}$
is a time at which a $\bar{Z}$ occurs
\item Each $\bar{H}$ occurs at a time $t_{\bar{H}}=t_{Z}+\tau$, where
$t_{Z}$ is a time at which a $Z$ occurs
\item Each $Z$ is connected to the previous $\bar{Z}$ by the equation
showing that the displacement of the trajectory between the two events
is $0$.
\item Each $\bar{Z}$ is connected to the previous $Z$ by the equation
showing that the displacement of the trajectory between the two events
is $0$.
\end{itemize}
As there is one equation for each symbol, we can solve the system
of equation to calculate the times. We then solve for the positions,
using the events to calculate the slope between the calculated times.
Finally, we generate a set of inequalities for each event:
\begin{itemize}
\item The time at which each event occurs must be consistent with the order
of sequence.
\item Each event occurring between consecutive $Z$ and $\bar{Z}$ events
occurs at $x>0$.
\item Each event occurring between consecutive $\bar{Z}$ and $Z$ events
occurs at $x<0$.
\end{itemize}
If the solved system of equations satisfies the set of inequalities
then the sequence is legal, for a given $\left(b,\tau\right)$. These
techniques allow us to use symbolic sequences to study solutions systematically.
Solving for the times and positions associated with events in a sequence,
we can plot the solution represented by the sequence. Additionally,
by changing the inequalities to equalities, we obtain the bifurcation
curves shown in Fig.~\ref{fig:bifurcations}.

\subsection{Calculation of the bifurcation curve of $\mathbb{P}$}

For $\tau>\frac{1}{2}$, $\mathbb{P}$ can written as
\begin{equation}
\mathbb{P}\left(S_{z}\right)=AS_{z}+B,
\end{equation}
where
\begin{equation}
B=\left(\begin{array}{c}
-\frac{1}{2}\\
-\frac{1}{2}\\
-\frac{1}{2}\\
-\frac{1}{2}\\
...\\
-\frac{1}{2}\\
\frac{b+2\tau(-1)^{n-1}}{b+(-1)^{n-1}}-\frac{1}{2}
\end{array}\right).
\end{equation}
Due to the sparsity of $A$, the characteristic equation can be readily
calculated as
\begin{equation}
(-1-\lambda)(-\lambda)^{n-1}+(-1)^{n-1}\frac{2(-1)^{n-1}}{b+(-1)^{n-1}}=0.
\end{equation}
This simplifies to
\begin{equation}
\lambda^{n}+\lambda^{n-1}-\frac{2(-1)^{n-1}}{b+(-1)^{n-1}}=0.
\end{equation}
By substituting $\lambda=e^{i\rho}$, we can solve for $b$ to determine
$b=b_{\text{bif }}$ for which the fixed point of $\mathbb{P}$ is
bifurcating as
\begin{equation}
e^{i\rho n}+e^{i\rho\left(n-1\right)}-\frac{2(-1)^{n-1}}{b_{\text{bif}}+(-1)^{n-1}}=0.\label{eq:a}
\end{equation}
Note that as $\frac{2(-1)^{n-1}}{b_{\text{bif}}+(-1)^{n-1}}\in\mathbb{R}$,
\begin{equation}
e^{i\rho n}=\overline{e^{i\rho\left(n-1\right)}}.\label{eq:b}
\end{equation}
So, $e^{i\rho\left(2n-1\right)}=1=e^{i2\pi k},k\in\mathbb{Z}$ , and
so $\rho=\frac{2\pi k}{2n-1}$. In order to satisfy Eq.~(\ref{eq:b}),
$k=n-1$. Substituting $\rho$ back into Eq.~(\ref{eq:a}),
\begin{equation}
\cos\left(\frac{2\pi n(n-1)}{2n-1}\right)+\cos\left(\frac{2\pi\left(n-1\right)^{2}}{2n-1}\right)-\frac{2(-1)^{n-1}}{b_{\text{bif}}+(-1)^{n-1}}=0,
\end{equation}
which simplifies under a sum-to-product cosine identity to
\begin{equation}
\cos\left(\frac{\pi\left(n-1\right)}{2n-1}\right)-\frac{1}{b_{\text{bif}}+(-1)^{n-1}}=0.
\end{equation}
Therefore,
\begin{equation}
b_{\text{bif}}(n)=\frac{1}{\cos\left(\frac{\pi\left(n-1\right)}{2n-1}\right)}-(-1)^{n-1},
\end{equation}
where $n=\left\lceil 2\tau\right\rceil $.

\bibliography{dde_paper}

\end{document}